# Forget BIT, It is All about TOKEN

## *— Towards Semantic Information Theory for LLMs*

Bo Bai

Despite the empirical successes of Large Language Models (LLMs), the prevailing paradigm is heuristic and experiment-driven, tethered to massive compute and data, while a first-principles theory remains absent. This treatise develops a **Semantic Information Theory** at the confluence of statistical physics, signal processing, and classical information theory, organized around a single paradigm shift: replacing the classical **BIT** — a microscopic substrate devoid of semantic content — with the macroscopic **TOKEN** as the atomic carrier of meaning and reasoning. Within this framework we recast attention and the Transformer as energy-based models, and interpret semantic embedding as vectorization on the semantic manifold. Modeling the LLM as a stateful channel with feedback, we adopt **Massey's directed information** as the native causal measure of autoregressive generation, from which we derive a **directed rate-distortion function** for pre-training, a **directed rate-reward function** for RL-based post-training, and a sub-martingale account of inference-time semantic information flow. This machinery makes precise the identification of next-token prediction with **Granger causal inference**, and sharpens the limits of LLM reasoning against **Pearl's Ladder of Causation** — affirming that **whereas the BIT defined the Information Epoch, the TOKEN will define the AI Epoch.**

**Dr. Bo Bai** (ee.borisbai@gmail.com) received his Ph.D. in Electronic Engineering from Tsinghua University, Beijing, in 2010, where he was recognized as a Young Academic Talent of Tsinghua University. He then served as a Research Associate in the Department of Electronic and Computer Engineering at the Hong Kong University of Science and Technology (2010–2012), and as an Assistant Professor in the Department of Electronic Engineering at Tsinghua University (2012–2017). He currently serves as Director of both the Leibniz Research Center and the Theory Lab, and Chief Scientist of Information Theory at 2012 Labs, Huawei Technologies Co., Ltd., Hong Kong. His research lies at the intersection of classical information theory, statistical physics, and graph theory, with a focus on their application to the foundations of LLMs, semantic communications, and data science.

An IEEE Senior Member, Dr. Bai has authored over 150 peer-reviewed publications in leading IEEE and ACM journals and flagship conferences, along with two book chapters and one textbook. He currently serves as an Associate Editor for *IEEE Communications Surveys & Tutorials*, and previously served as the founding Vice Chair of the IEEE TCCN Special Interest Group on Social-Behavior-Driven Cognitive Radio Networks. He has contributed to the Technical Program Committees of numerous IEEE conferences and regularly reviews for leading IEEE and ACM transactions and journals. He received the 2016 IEEE ICC Best Paper Award.

**Acknowledgments**: I am indebted to my colleagues Tingyi Wu, Xueyan Niu, Ken Zhang, Chaorui Zhang, Yuan Lan, Zhimeng Zhong, Bing Chen, and Qunying Zhang for the many stimulating discussions that shaped the ideas in this treatise. I also thank the broader Theory Lab community for fostering an intellectual environment in which cross-disciplinary inquiry — at the confluence of information theory, statistical physics, and modern machine learning — can flourish.

*"The interesting question is,*
*just because something thinks differently from you,*
*does that mean it's not thinking?"*

— Alan Turing as portrayed in *The Imitation Game*, 2014

# Contents

# 1. Introduction

The late 2022 release of ChatGPT marked a watershed moment, astonishing the world with its capabilities and catalyzing a global paradigm shift across research, industry, and public discourse. Several months later, we had the privilege of hosting Prof. Erdal Arikan — inventor of Polar codes — for a panel colloquium.[1] In the opening exchange, my colleague Dr. Wu asked: "Prof. Arikan, what do you consider the paramount conceptual breakthrough of the Information Epoch?" After a brief pause, the professor answered without hesitation: "The **BIT**. I believe the bit stands as the greatest invention of the Information Epoch." The remark lingered, precipitating a deeper question: in the wake of ChatGPT's emergence, what is the corresponding atomic concept of the AI Epoch — one possessing the same foundational primacy as the bit? After years of sustained engagement with Large Language Model (LLM) research, the answer, to us, has become unambiguous: that foundational construct is the **TOKEN**.

Drawing inspiration from Claude Shannon's seminal 1948 paper [1], this work seeks to build the theoretical foundations of LLMs from an inferential standpoint. Shannon's starting point was the pursuit of reliable information transmission over noisy channels. From this pragmatic engineering origin, he erected the mathematical edifice now universally known as **Information Theory**, a rigorous and complete taxonomy of concepts and limit theorems. In 1949, Warren Weaver and Claude Shannon co-authored an exposition that partitioned the overarching communication problem into three hierarchical tiers [2]:

- **Level A (The Technical Problem):** How accurately can the symbols of communication be transmitted?
- **Level B (The Semantic Problem):** How precisely do the transmitted symbols convey the intended meaning?
- **Level C (The Effectiveness Problem):** How effectively does the received meaning affect conduct in the desired way?

Shannon himself acknowledged that his framework was deliberately confined to Level A: the problem of reliable communication. Within his paradigm, information is equated with statistical uncertainty, and the theory is agnostic to the semantic meaning and pragmatic significance of the messages it transmits, concerning itself solely with the error-free reception of their binary representations. The central thesis of this work is that by pivoting Shannon's apparatus from bit to token — that is, by lifting the atomic unit from a semantically inert symbol to a semantically charged one — the mathematical and inferential mechanics of LLMs can be systematically elucidated, opening a principled path from Level A into Levels B and C. We designate this generalized, interdisciplinary framework as **Semantic Information Theory**.

The formal study of semantics traces back to Rudolf Carnap, whose analyses spanned empiricism, ontology, linguistics, and logic [3], [4], [5], and whose treatise [6] offered a systematic exposition of semantics and modal logic; contemporary trajectories stemming from these foundations are surveyed in [7], [8].

[1]The event was broadcast live through the Chaspark website (https://www.chaspark.com) and became the most viewed live event of 2023. Video: https://www.chaspark.com/#/live/862749892264718336?anchorV=1022651139410739200

Building on Carnap's epistemological framework, Ray Solomonoff introduced **algorithmic probability** and embedded it within the Bayesian paradigm to forge a formal theory of inductive inference [9], [10], [11]: the prior probability of a sequence is dictated by its algorithmic complexity, with the most succinct generative program receiving the greatest weight and its length defining the sequence's **Kolmogorov complexity**. Although Shannon's statistical information theory [1] was already established by 1948, Kolmogorov complexity was formalized roughly two decades later [12], [13] — not as a historical precursor to Shannon's apparatus, but as a retrospective **algorithmic** reformulation of it, recasting information content as the length of the shortest generative program on a universal Turing machine. Under mild regularity conditions, Shannon entropy and expected Kolmogorov complexity coincide up to an additive constant, so the two frameworks converge asymptotically while offering complementary epistemic stances: statistical on the one hand, Turing-complete and generative on the other.

Together, the Solomonoff prior and Kolmogorov complexity underpin **universal reinforcement learning** for sequential decision-making in artificial agents [14]. Yet Kolmogorov complexity is exactly Turing-uncomputable, rendering both the Kolmogorov and Solomonoff edifices mathematically elegant but fundamentally intractable for empirical AI — a tension that motivates the probabilistic, estimator-based approach taken in this work.

When Kolmogorov complexity is evaluated over the sample sequences of a random variable, its expectation coincides exactly with the Shannon entropy [15], [16]; Shannon's information theory is thus often viewed as a probabilistic instantiation of Kolmogorov's algorithmic theory. Yet it is the probabilistic framework that has proven decisively more instrumental for modern neural networks and LLMs, for a simple reason: the core information-theoretic quantities — entropy, mutual information, and Kullback-Leibler (KL) divergence — are empirically computable, admitting tractable estimators at scale [17]. This computability is precisely what the uncomputability of Kolmogorov complexity denied its algorithmic counterpart, and it is the pragmatic pivot that resonates with Richard Sutton's *Bitter Lesson* [18]: "The biggest lesson that can be read from 70 years of AI research is that general methods that leverage computation are ultimately the most effective, and by a large margin."

The pivotal challenge in re-centring Shannon's theory around the token is to devise a computable and mathematically tractable representation of semantics. Classical source coding optimizes for the minimal binary compression of messages and deliberately abstracts away their intrinsic semantic content; what is needed instead is a representation in which semantic structure is itself the object of computation. The use of **vectors** as information carriers traces back to Luhn [19], and vector-space models became the bedrock of information retrieval [20]. Bengio et al. then took the decisive step of jointly learning low-dimensional distributed word representations — **word embeddings** — alongside language models [21], integrating vector semantics with neural architectures for the first time. Mikolov et al. sharpened this paradigm with the CBOW and Skip-gram architectures, showing that high-fidelity word vectors could be trained on massive corpora with remarkably simple neural networks [22], and further revealed that these dense vectors encode linear substructures that reflect meaningful semantic relationships algebraically [23]. This catalyzed an explosion of embedding research, culminating in landmark frameworks such as GloVe [24], FastText [25], and ELMo [26], and establishing vector semantics as the indispensable computational substrate for modern natural language processing (NLP) and LLMs [27].

Isolated vector representations, however, capture only atomic, token-level semantics. Scaling semantic representation and generation to coherent sentences, paragraphs, and full documents long stood as a computational and theoretical bottleneck. The advent of the **Transformer** architecture [28], built around the **Attention** mechanism, resolved this bottleneck and unleashed unprecedented representational power across the NLP spectrum. On this foundation, OpenAI introduced the GPT lineage, whose successive generations exhibited remarkable generalization across an ever-expanding range of complex tasks [29], [30], [31], [32]. More recently, DeepSeek has contributed architectural refinements that sharply improve training efficiency while preserving the core Transformer topology, yielding LLMs of notable inferential capability [33], [34]. Yet a mathematically unified theoretical account of the Transformer mechanism remains elusive, so contemporary architectural progress is still overwhelmingly driven by heuristic, large-scale GPU empiricism — a trajectory whose computational costs are rapidly becoming unsustainable.

To transcend this empirical bottleneck, this work establishes semantic information theory as an interdisciplinary theoretical framework for decoding the “black box” of LLMs, integrating analytical machineries from statistical physics, signal processing, and information theory — all re-centred around the token.

From a **statistical physics** perspective, we draw a theoretical linkage between LLMs and classical **energy-based models (EBMs)**: the Attention mechanism and feed-forward networks (FFNs) of the Transformer are formalized as continuous-state Hopfield networks. This physical framing supports a generalization-error bound via local Rademacher complexity [35], [36], a memory-capacity analysis through a generalized Gardner capacity that establishes the Transformer’s theoretical capability for exponential knowledge storage [37], [38], [39], and a characterization of inferential limits through Granger causality, from which we argue that next-token prediction ultimately rests on temporal statistical inference rather than genuine logical reasoning [40], [41].

From a **signal processing** perspective, we study the token-level semantic vector space and its dimensional compression. Semantic distortion is quantified via the Gromov-Wasserstein distance [42], [43], and semantic compression is formalized through the Johnson-Lindenstrauss (JL) lemma. On this geometric foundation, we formulate an information-theoretically optimal semantic vectorization scheme for next-token prediction and expose its theoretical isomorphism to contrastive predictive coding (CPC) [44], [45]. Anchored in the theory of non-linear time-varying **vector autoregression (VAR)** processes [46], we give a rigorous mathematical definition of autoregressive LLMs and show that the Transformer [28] is a specialized instantiation of this generalized VAR formulation — a vantage that simultaneously exposes the structural constraints of emerging linear architectures such as Mamba/Mamba2 [47], [48] and LLaDA [49]. Finally, through variational inference we derive the evidence lower bound (ELBO) of the Transformer across both training and inference phases [50].

From an **information theoretic** perspective, we build a probabilistic model of the LLM as a next-token predictor and expose its mathematical isomorphism to a discrete-time commu-nication channel with feedback and state. This prompts a decisive departure from classical Shannon tenets: **the channel is reconceptualized from a passive conduit for reliable information transmission into an active generative engine, with the overarching objective shifting from exact sequence recovery to the generation of sequences that satisfy prescribed semantic and pragmatic constraints**. On this premise, we introduce the **directed rate-distortion function** as a unifying theoretical metric for the LLM pre-training phase [51], [52], and the **directed rate-reward function** as its counterpart for the reinforcement-learning-driven post-training phase [53], from which we argue that LLMs asymptotically approximate human-level Granger causality in next-token prediction [40]. Finally, we axiomatize the inference-phase dynamics by analyzing semantic information flow through sub-martingales, and synthesize a set of semantic information-theoretic measures for LLMs grounded in directed-information estimation.

The remainder of this paper is organized as follows. Section 2 develops the statistical-physics perspective, casting the Attention and FFN modules as EBMs and thereby deriving a local-Rademacher generalization bound, a Gardner-type memory capacity, and a Granger-causal characterization of the Transformer's inferential limits. Section 3 adopts a signal-processing lens: it scrutinizes the geometry of the token-level semantic vector space, recasts the Transformer as a non-linear time-varying VAR process, establishes the ELBO governing its representation, and situates emerging alternative architectures within the same formal substrate. Section 4 then re-centres Shannon's apparatus around the token, introducing directed-information-based measures for LLMs that span the pre-training, post-training, and inference phases, and articulating the epistemological duality between Granger- and Pearl-causal readings of next-token prediction. Section 5 closes the paper with a synthesis of open problems and outlooks at the intersection of semantic information theory and LLMs.

# 2. Statistical Physics Perspective

## 2.1. Neural Networks and Statistical Physics

The 2024 Nobel Prize in Physics was awarded to John Hopfield and Geoffrey Hinton, with the citation: *For foundational discoveries and inventions that enable machine learning with artificial neural networks.* The decision initially drew some public bewilderment, with critics suggesting that the Nobel Committee had merely capitulated to the AI zeitgeist. Such readings, however, miss the deep and long-standing isomorphism between neural networks and statistical mechanics — an intellectual lineage woven into the very origins of both disciplines.

In 1982, John Hopfield introduced the **Hopfield network** — an architecture with genuine associative-memory capabilities that electrified the scientific community [54], reinvigorated global interest in neural networks, and is widely regarded as a decisive catalyst for AI's emergence from its long winter. Amit et al. soon imported the spin-glass formalism from statistical mechanics to rigorously model and analyze the Hopfield network [55]. Geoffrey Hinton — later dubbed the "Godfather of AI" — was among the first computer scientists to recognize the explanatory power of statistical physics for neural architectures: in 1985, he and his collaborators proposed the **Boltzmann machine**, installing the EBM as its theoretical bedrock [56], [57].

Beyond the Nobel laureates above, physicist Elizabeth Gardner made equally indispensable contributions to this lineage. In a remarkable burst of productivity in 1988, she published three treatises that systematically characterized the storage capacity of the Hopfield network, bounding the maximum number of random binary patterns it could reliably encode [37], [38], [39]. This theoretical limit is now enshrined as the **Gardner capacity**, conventionally denoted $\alpha_c$. Her analysis leaned heavily on the spin-glass model and the **replica method** — a technique pioneered by Giorgio Parisi, recipient of the 2021 Nobel Prize in Physics [58], [59]. The author had the privilege of interviewing Prof. Parisi in 2025, in a wide-ranging conversation on the synergy between AI and statistical physics.[2]

Because the spin variables are strictly constrained to the binary domain $\{-1, +1\}$, the classical Hopfield network and its Gardner capacity are defined at the **BIT** level — a fact that forges an intrinsic isomorphism with Shannon's information theory. Nicolas Sourlas formalized this synergy in 1989, establishing that the retrieval dynamics of the Hopfield model are mathematically equivalent to the decoding of error-correcting codes [60]. Under this mapping, stored memory patterns correspond to source messages, Hebbian learning rules to the generation of redundant codewords, synaptic noise to channel interference, and associative retrieval to error-correcting decoding. Consequently, the Shannon channel capacity manifests as the Gardner capacity, serving as the absolute theoretical upper bound for associative memory storage. This fusion subsequently catalyzed the widespread deployment of statistical-mechanics methods in modern coding theory, shaping the development of LDPC codes, Turbo codes, sparse regression codes (SRC), and beyond [61], [62].

Despite this mathematical correspondence with Shannon's information theory, the strict binary formulation of the classical Hopfield network severely constrains its applicability to the semantic complexities of NLP. The modern renaissance of Hopfield architectures was catalyzed by Krotov and Hopfield in 2016, who generalized the standard pairwise interaction $F(x) = x^2$ to higher-order many-body interactions $F(x) = x^n$ and showed that the storage capacity scales from $\alpha_c \sim \mathcal{O}(N)$ to $\alpha_c \sim \mathcal{O}(N^{n-1})$ as the interaction order $n$ grows [63]. Demircigil et al. then showed that an exponential interaction $F(x) = \exp(x)$ unlocks an exponential storage capacity $\alpha_c \sim \mathcal{O}(\exp(\eta N))$ for a constant $\eta > 0$ [64]. Building on these generalizations, Ramsauer et al. formulated a **continuous-state Hopfield network** and proved that its update dynamics are mathematically isomorphic to the Attention mechanism of the Transformer [65]. Concurrently, Krotov and Hopfield introduced a unified continuous-state Lagrangian bridging discrete and continuous states, as well as auto-associative and hetero-associative memory [66]. Pushing the unification further, Hoover et al. constructed the energy Transformer, recasting sequence processing as a trajectory of energy minimization over a large continuous Hopfield network and subsuming both the Attention and FFN modules within a single energy functional [67]. Most recently, Nicoletti et al. carried out a comprehensive replica analysis of vector-spin Hopfield networks, characterizing the spin-glass phases and replica symmetry breaking (RSB) that emerge near the saturation threshold [68].

[2]Video: https://www.chaspark.com/#/hotspots/1205217438115655680°

Taken together, this trajectory reveals a deep, symbiotic lineage between statistical physics and neural architectures, most visibly embodied by the evolution of Hopfield networks. The decisive step — from the classical binary Hopfield paradigm to its modern continuous-state incarnations — was catalyzed by the systematic algebraic generalization of the underlying energy functions. Conceptually, this is a passage from the **BIT** to the **TOKEN**: it supplies the theoretical apparatus through which statistical physics can now decipher the inner mechanisms of the Transformer, and it precipitates the broader paradigm shift from classical **Shannon Information Theory** to **Semantic Information Theory**.

## 2.2. EBM Formulation of the Attention Module

There is now broad consensus in the AI community that the core operational objective of LLMs is **next-token prediction**, and that the Transformer has become the dominant architecture for realizing it. As illustrated in Figure 1, the analysis that follows is restricted to a decoder-only Transformer, whose computation decomposes into two canonical modules: the **Attention module** and the **FFN module** [28].

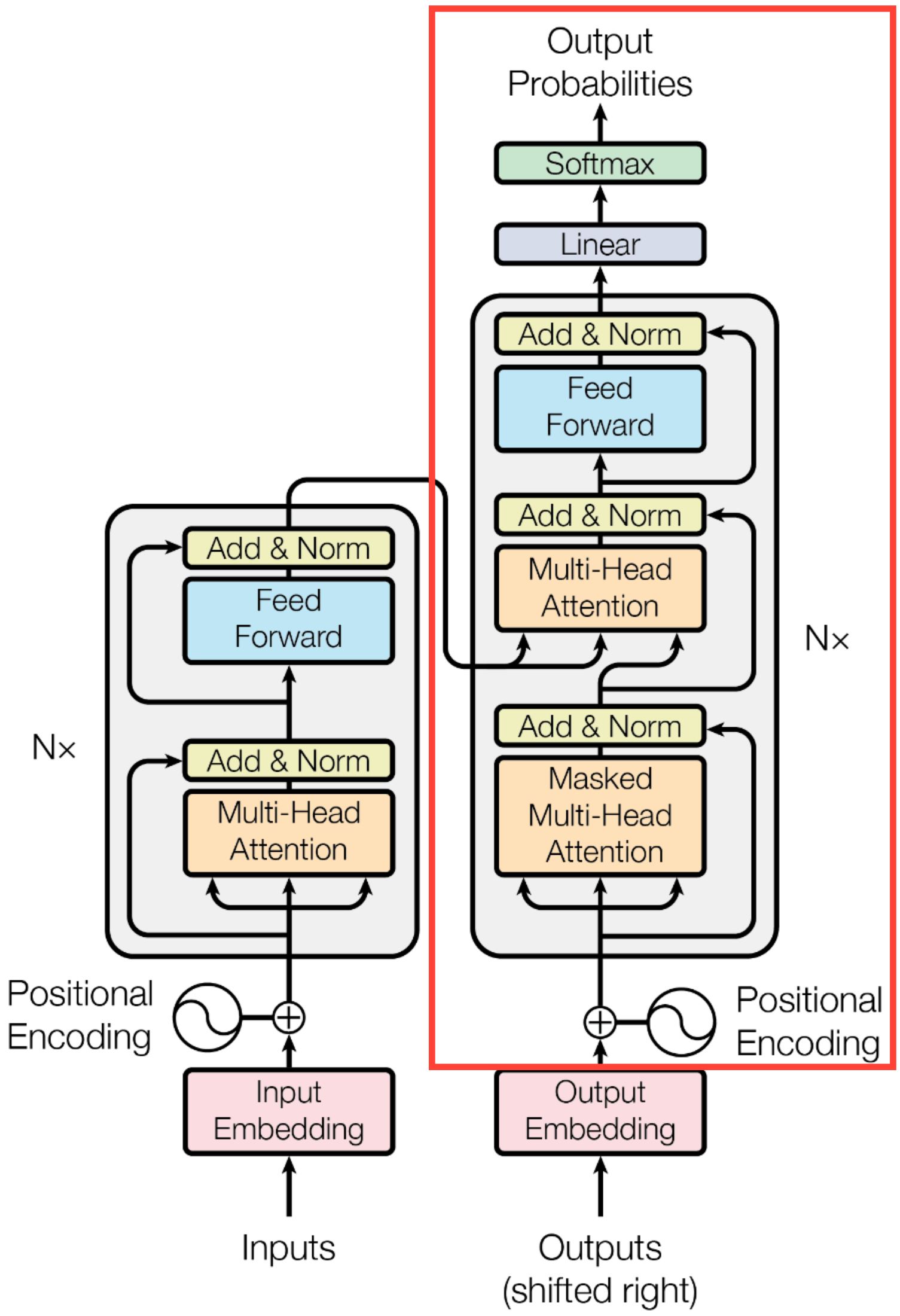

Figure 1: Schematic of the Transformer architecture, with the decoder-only path highlighted in the red box.

For a given Transformer, let $\mathcal{U}$ denote the space of all admissible tokens with cardinality $|\mathcal{U}| = M$. This discrete set is called the **vocabulary** in machine-learning terminology and coincides with the **codebook** of classical information theory. Consider an input prompt of length $n$, represented by a sequence $\boldsymbol{s}_{1:n}$ with $\boldsymbol{s}_i \in \mathcal{U}$ for $i = 1, \ldots, n$; the autoregressive objective is then to predict the subsequent tokens $\boldsymbol{u}_i \in \mathcal{U}$ for $i = n+1, \ldots, N$. For notational uniformity, we set $\boldsymbol{u}_i = \boldsymbol{s}_i$ for $i = 1, \ldots, n$. The contextualized output of the Attention mechanism is

$$\tilde{\boldsymbol{v}}_i = \sum_{j=1}^{i-1} \pi_{ij} \boldsymbol{W}_{\mathrm{v}} \boldsymbol{u}_j, \quad i = n+1, \ldots, N, \tag{1}$$

where $\pi_{ij}$ denotes the normalized Attention weight and $\boldsymbol{v}_j = \boldsymbol{W}_{\mathrm{v}} \boldsymbol{u}_j$ is the **value vector**. The resulting $\tilde{\boldsymbol{v}}_i$ acts as a contextual aggregator, distilling from the prefix $\boldsymbol{u}_{1:i-1}$ all historically relevant semantic information needed to predict $\boldsymbol{u}_i$. Expanding via the softmax definition, the Attention weights admit the closed form

$$\pi_{ij} = \frac{\exp\big(\langle \boldsymbol{W}_{\mathrm{q}} \boldsymbol{u}_{i-1}, \boldsymbol{W}_{\mathrm{k}} \boldsymbol{u}_j \rangle\big)}{\sum_{l=1}^{i-1} \exp\big(\langle \boldsymbol{W}_{\mathrm{q}} \boldsymbol{u}_{i-1}, \boldsymbol{W}_{\mathrm{k}} \boldsymbol{u}_l \rangle\big)} = \frac{\exp\big(\langle \boldsymbol{q}_{i-1}, \boldsymbol{k}_j \rangle\big)}{\sum_{l=1}^{i-1} \exp\big(\langle \boldsymbol{q}_{i-1}, \boldsymbol{k}_l \rangle\big)}, \quad j = 1, ..., i-1, \tag{2}$$

where $\boldsymbol{q}_{i-1} = \boldsymbol{W}_{\mathrm{q}} \boldsymbol{u}_{i-1}$ is the **query vector** and $\boldsymbol{k}_j = \boldsymbol{W}_{\mathrm{k}} \boldsymbol{u}_j$ the **key vector**. Their inner product $\langle \boldsymbol{q}_{i-1}, \boldsymbol{k}_j \rangle$ quantifies the semantic congruency between the historical context $\boldsymbol{u}_j$ and the immediate predecessor $\boldsymbol{u}_{i-1}$ when forecasting the next token $\boldsymbol{u}_i$.

Defining the interaction matrix $\boldsymbol{J} = \boldsymbol{W}_{\mathrm{q}}^T \boldsymbol{W}_{\mathrm{k}}$, the inner product rewrites as

$$\langle \boldsymbol{u}_{i-1}, \boldsymbol{J} \boldsymbol{u}_j \rangle, \tag{3}$$

which is precisely a **bilinear form** — the canonical geometric device for modeling asymmetric relationships through inner products. The Attention mechanism is thus intrinsically structured to encode the **asymmetric semantic relevance** between the current token and its historical antecedents, providing the predictive substrate for autoregressive generation.

The softmax operator governing $\pi_{ij}$ plays a deeper theoretical role than it first appears. By **Jaynes' maximum entropy principle**, the distribution produced by softmax is precisely the maximum-entropy discrete distribution subject to empirical mean constraints — equivalently, to bounds on the average activations of the network [69]. In inferential terms, the softmax output is the most unbiased statistical estimate attainable under maximal epistemic uncertainty.

In Hinton's paradigm, we can now formalize the EBM of the Attention mechanism through the following definition.

**Definition 2.2.1** (EBM Formulation of the Attention Mechanism): Given the query token $\boldsymbol{u}_{i-1}$ and the interaction matrix $\boldsymbol{J} = \boldsymbol{W}_{\mathrm{q}}^T \boldsymbol{W}_{\mathrm{k}}$, the localized energy of a prefix token $\boldsymbol{u}_j$ at position $j \in \{1, ..., i-1\}$ is

$$E_{\mathrm{A}}\big(\boldsymbol{u}_j; \boldsymbol{u}_{i-1}, \boldsymbol{J}\big) = -\langle \boldsymbol{u}_{i-1}, \boldsymbol{J} \boldsymbol{u}_j \rangle. \tag{4}$$

The induced Boltzmann distribution over prefix positions, which coincides with the Attention weights $\pi_{ij}$, is

$$\mathbb{P}_{\mathrm{A}}\big(\boldsymbol{u}_j; \boldsymbol{u}_{i-1}, \boldsymbol{J}\big) = \frac{1}{Z_{\mathrm{A}}} \exp\big(-E_{\mathrm{A}}\big(\boldsymbol{u}_j; \boldsymbol{u}_{i-1}, \boldsymbol{J}\big)\big), \quad j = 1, ..., i-1, \tag{5}$$

with the partition function summed over the historical context

$$Z_{\mathrm{A}} = \sum_{j=1}^{i-1} \exp\big(-E_{\mathrm{A}}\big(\boldsymbol{u}_j; \boldsymbol{u}_{i-1}, \boldsymbol{J}\big)\big). \tag{6}$$

By mathematical construction, it naturally holds that $\mathbb{P}_{\mathrm{A}}\big(\boldsymbol{u}_j\big) = \pi_{ij}$ for all $j = 1, ..., i-1$. Viewed through this statistical mechanics prism, the fundamental optimization objective of the Attention mechanism is to learn a parametric configuration $\boldsymbol{J}$ that systematically maps the highest asymmetric semantic relevance to the lowest thermodynamic energy state. This energy-minimization logic profoundly dovetails with the principles of variational inference predicated upon the latent variable $J$, a theoretical nexus we shall rigorously dissect in Section 3.3.

Within the vanguard of statistical physics analyses applied to the Attention mechanism, the academic community has recently formalized several pivotal theoretical breakthroughs, systematically synthesized as follows:

- **Attention dynamics and phase collapse:** Poc-López et al. deployed **dynamical mean-field theory (DMFT)** — a technique closely allied with the replica method — to map the forward propagation and backward updates of tokens across multiple Transformer layers onto a macroscopic interacting-particle system [70]. They showed that unmediated stacking of pure self-Attention layers inevitably forces tokens at all positions to collapse into a degenerate, uniform representation as depth grows. Crucially, the analysis supplies a physical rationale for why state-of-the-art Transformers require residual connections, layer normalization, and carefully calibrated weight-initialization scaling: these are not mere engineering heuristics but thermodynamic interventions that break the uniform phase emerging in deep strata, preserving the spatial anisotropy and representational entropy of token embeddings in high-dimensional semantic space.
- **Positional coding and semantic phase transitions:** Cui et al. developed a rigorous theoretical framework for dot-product Attention that uncovers the physical mechanism by which its dynamics undergo a sharply nonlinear, discontinuous transition from a position-dominated regime to a semantics-driven one [71]. The resulting picture casts the sudden emergence of in-context understanding in LLMs as the complex system crossing a thermodynamic critical point — a macroscopic **phase transition** from a positional phase to a semantic phase.
- **Linear Attention and spin-glass mappings:** Linear Attention architectures such as Mamba/Mamba2 [47], [48] sharply reduce computational complexity but suffer from a fundamental deficiency: they cannot faithfully encode the asymmetric semantic dependencies between tokens. To expose the source of this limitation, Li et al. established a rigorous mapping between the linear-Attention Transformer and a real-valued **spin-glass** system, in which the statistical disorder of the training data plays the role of coupling constants and external quenched fields [72]. Their analysis shows that scaling task diversity forces the underlying Boltzmann distribution to collapse onto a unique, optimal weight configuration — a critical phenomenon mathematically equivalent to a thermodynamic phase transition. Pure linear Attention is therefore theoretically expressively limited, which helps rationalize the industry's recent pivot to hybrid Attention architectures such as Qwen2.5-Math and Qwen3-Next.
- **Sparse Attention and retrieval dynamics:** Reducing computational complexity while preserving the capacity to model asymmetric semantic dependencies remains a central architectural challenge, and sparse Attention mechanisms have emerged as a highly active research frontier. Most notably, DeepSeek-V3.2 has put into production a carefully engineered variant, **DeepSeek sparse Attention (DSA)**, with compelling empirical efficacy [73]. On the theoretical side, Jerry et al. extended the modern Hopfield formalism by replacing the canonical log-sum-exp energy functional with an $\alpha$-entmax one, yielding retrieval dynamics that are mathematically equivalent to sparse Attention and that suppress the thermodynamic interference of irrelevant memory patterns [74]. Formally, designing an optimal sparse Attention mask amounts to solving, at each generation step $i = n + 1, ..., N$, the $\ell_0$-minimization

$$\begin{aligned} \min_{\pi'_{i,1:i-1}} \quad & \left\| \pi'_{i,1:i-1} \right\|_0 \\ \text{s.t.} \quad & \mathbb{E}_{\boldsymbol{u}_{1:i-1}} \left[ \left\| \sum_{j=1}^{i-1} \left( \pi_{ij} - \pi'_{ij} \right) \boldsymbol{W}_{\mathrm{v}} \boldsymbol{u}_j \right\|_2^2 \right] \leq \varepsilon, \end{aligned} \tag{7}$$

where $\varepsilon > 0$ is a prescribed fidelity tolerance, the expectation is over the data distribution of prefixes $\boldsymbol{u}_{1:i-1}$, and $\pi_{ij}, \pi'_{ij} \in [0, 1]$ denote respectively the dense and the sparsified Attention weights at position $j$, with both $\pi_{i,1:i-1}$ and $\pi'_{i,1:i-1}$ normalized to probability vectors over $(i - 1)$ positions.

- **Multi-head Attention and the replica method:** Through the lens of statistical physics, multi-head Attention can be read as a macroscopic realization of the replica method. Bordelon et al. deployed DMFT to analyze the training dynamics of multi-head Transformers in several asymptotic thermodynamic limits: infinite head count, infinite key/query dimensionality, and infinite depth. Their central finding is that the choice of parameter scaling — particularly whether the Attention-layer output is normalized by the head count — induces qualitative bifurcations in macroscopic behavior: in some regimes the Attention heads evolve as independent replicas, whereas in others pronounced inter-head correlations emerge [75].

## 2.3. EBM of the Transformer Architecture

Building on the localized EBM of the Attention mechanism, we now derive an EBM for the complete Transformer. This global thermodynamic formalization supplies the theoretical foundation needed to dissect the macroscopic properties of the architecture — in particular, its generalization error, memory capacity, and inferential limits. In what follows, the FFN module is represented by a non-linear mapping $\Psi_{\mathrm{FFN}}$ acting on the Attention output $\tilde{\boldsymbol{v}}_i$, with $\mathcal{W}$ and $\mathcal{B}$ denoting the sets of weight matrices and bias vectors across all layers; we assemble the full parameter configuration as $\mathcal{P} = \left\{ \boldsymbol{W}_{\mathrm{q}}, \boldsymbol{W}_{\mathrm{k}}, \boldsymbol{W}_{\mathrm{v}}, \mathcal{W}, \mathcal{B} \right\}$, and write the discrete vocabulary as $\mathcal{U} = \{\tilde{\boldsymbol{u}}_1, ..., \tilde{\boldsymbol{u}}_M\}$. Let $T > 0$ denote the thermodynamic temperature. With these conventions, the global EBM of the Transformer is as follows.

**Definition 2.3.1** (EBM of the Transformer Architecture): For a candidate token $\tilde{\boldsymbol{u}}_m \in \mathcal{U}$ and parameter configuration $\mathcal{P}$, at generation step $i \in \{n+1, ..., N\}$, the global energy is

$$E_{\mathrm{T}}(\tilde{\boldsymbol{u}}_m; \mathcal{P}) = -\langle \tilde{\boldsymbol{u}}_m, \Psi_{\mathrm{FFN}}(\tilde{\boldsymbol{v}}_i) \rangle, \tag{8}$$

where $\tilde{\boldsymbol{v}}_i$ is the Attention output at step $i$ defined via Definition 2.2.1. The corresponding Boltzmann distribution over the vocabulary is

$$\mathbb{P}_{\mathrm{T}}(\tilde{\boldsymbol{u}}_m; \mathcal{P}) = \frac{1}{Z_{\mathrm{T}}} \exp\left(-\frac{1}{T} E_{\mathrm{T}}(\tilde{\boldsymbol{u}}_m; \mathcal{P})\right), \quad m = 1, ..., M, \tag{9}$$

with the partition function summed over the vocabulary

$$Z_{\mathrm{T}} = \sum_{m=1}^{M} \exp\left(-\frac{1}{T} E_{\mathrm{T}}(\tilde{\boldsymbol{u}}_m; \mathcal{P})\right). \tag{10}$$

Observe that the vector $\boldsymbol{z}_i(\mathcal{P}) = -\frac{1}{T}(E_{\mathrm{T}}(\tilde{\boldsymbol{u}}_1; \mathcal{P}), ..., E_{\mathrm{T}}(\tilde{\boldsymbol{u}}_M; \mathcal{P}))$ is precisely the pre-softmax logits vector over the vocabulary at generation step $i \in \{n+1, ..., N\}$; the dependence of $\boldsymbol{z}_i$ on $i$ enters through the Attention output $\tilde{\boldsymbol{v}}_i$ embedded in $E_{\mathrm{T}}$. At inference time, the empirical output token $\boldsymbol{u}_i$ is typically drawn by stochastic sampling heuristics (temperature scaling, top-$k$ selection, and the like); the inferential limit — attained as $T \to 0$ — coincides with the token state that minimizes the global energy, or equivalently maximizes the Boltzmann probability. Dually, the training objective is to converge on a parameter configuration $\mathcal{P}$ that minimizes the expected energy over the empirical training distribution. For notational precision, we write $\mathcal{P}^{\dagger}$ for the converged parameters after pre-training, and $\mathcal{P}^{\ddagger}$ for the refined parameters produced by post-training alignment (e.g., RLHF).

### 2.3.1. Generalization Error Bound

Generalization error is the canonical metric for the out-of-sample inferential performance of LLMs. Building on the global EBM of the Transformer, we derive a rigorous upper bound on this error by combining two complementary tools from [35]:

- **Local Rademacher complexity** refines the classical global Rademacher complexity by restricting the supremum to function sub-classes with bounded variance, producing substantially tighter generalization bounds through a sub-root fixed-point characterization.
- **The Bernstein condition for non-negative bounded losses** couples the variance of the loss to its mean via $\mathrm{Var}(f) \leq b \cdot \mathbb{E}[f]$ for an absolute constant $b > 0$; combined with the local-Rademacher fixed point, it converts the classical slow rate $O\left(1/\sqrt{K}\right)$ into the fast rate $O(1/K)$.

**Cross-entropy** is the canonical loss for training LLMs; at inference time it quantifies the statistical alignment between the model's predictive distribution and the ground-truth human target. Fix a deterministic prefix $\boldsymbol{s}_{1:n}$. Let the random variable $U_i$ denote the $i$-th token synthesized by the LLM during autoregressive inference, and $U_i^{\mathrm{h}}$ the optimal target token mandated by human preference at the same step. For $i = n + 1, ..., N$, the fully aligned (post-trained) LLM induces the conditional distribution

$$\mathbb{P}_i(\mathcal{P}^\ddagger) = \mathbb{P}(U_i \mid U^{\mathrm{h}}_{1:i-1}; \mathcal{P}^\ddagger) \tag{11}$$

over the vocabulary $\mathcal{U}$, conditioned on the human-aligned prefix $U^{\mathrm{h}}_{1:i-1}$; this is exactly the Transformer Boltzmann distribution $\mathbb{P}_{\mathrm{T}}(\cdot; \mathcal{P}^\ddagger)$ of Definition 2.3.1, evaluated with the Attention output $\tilde{\boldsymbol{v}}_i$ induced by that prefix. In parallel,

$$\mathbb{P}^{\mathrm{h}}_i = \mathbb{P}(U^{\mathrm{h}}_i \mid U^{\mathrm{h}}_{1:i-1}) \tag{12}$$

denotes the ground-truth conditional distribution of the human target. Empirical realizations of $U_i$ and $U^{\mathrm{h}}_i$ are written as $\boldsymbol{u}_i$ and $\boldsymbol{u}^{\mathrm{h}}_i$; in the low-temperature limit $T \to 0$, the sampled $\boldsymbol{u}_i$ converges to the minimizer of the global energy — equivalently, the maximizer of $\mathbb{P}_i(\mathcal{P}^\ddagger)$. The generalization error is then the cross-entropy

$$H(\mathbb{P}^{\mathrm{h}}_i \wedge \mathbb{P}_i(\mathcal{P}^\ddagger)) = -\mathbb{E}_{\mathbb{P}^{\mathrm{h}}_i}[\log \mathbb{P}_i(\mathcal{P}^\ddagger)] = \sum_{m=1}^{M} \mathbb{P}^{\mathrm{h}}_i(\tilde{\boldsymbol{u}}_m) \log \frac{1}{\mathbb{P}_i(\tilde{\boldsymbol{u}}_m; \mathcal{P}^\ddagger)}. \tag{13}$$

An absolute upper bound on this cross-entropy is established by the following theorem.

**Theorem 2.3.1.1** (Generalization Error Bound of the Transformer): Let $\Theta^{\ddagger}$ denote a compact admissible post-training parameter class, frozen prior to inference (hence independent of the inference-time samples introduced below), and let

$$d := \operatorname{pdim}\left(\left\{E_{\mathrm{T}}(\cdot;\theta) : \theta \in \Theta^{\ddagger}\right\}\right) < \infty \tag{14}$$

denote the pseudo-dimension of the induced energy function class. Assume $\Theta^{\ddagger}$ contains the converged configuration $\mathcal{P}^{\ddagger}$, and further assume: (i) the energy functional is uniformly bounded, $|E_{\mathrm{T}}(\tilde{\boldsymbol{u}}_m;\theta)| \leq B$ for all $\tilde{\boldsymbol{u}}_m \in \mathcal{U}$ and all $\theta \in \Theta^{\ddagger}$; and (ii) the vocabulary is sufficiently large relative to the thermodynamic temperature, $\log M \geq 3B/T$. Define the loss range constant

$$G := \log M + 2B/T. \tag{15}$$

For each inference step $i = n+1, ..., N$, conditioned on the ground-truth human prefix $\boldsymbol{u}^{\mathrm{h}}_{1:i-1}$, draw $K$ i.i.d. empirical samples $\left\{\boldsymbol{u}^{\mathrm{h}}_{i,k}\right\}_{k=1}^{K}$ and let $m^{\mathrm{h}}_{i,k} \in \{1, ..., M\}$ denote the unique vocabulary index satisfying $\boldsymbol{u}^{\mathrm{h}}_{i,k} = \tilde{\boldsymbol{u}}_{m^{\mathrm{h}}_{i,k}}$. Then, for any prescribed confidence parameter $\delta \in (0,1)$, the generalization error of the Transformer architecture is bounded from above uniformly across all inference steps $i = n+1, ..., N$ with probability at least $1-\delta$:

$$H\left(\mathbb{P}^{\mathrm{h}}_i \wedge \mathbb{P}_i\left(\mathcal{P}^{\ddagger}\right)\right) \leq 2\hat{\mathcal{L}}_i + cG\frac{d + \log(2(N-n)/\delta)}{K}, \tag{16}$$

where $\hat{\mathcal{L}}_i = \frac{1}{K}\sum_{k=1}^{K} \log 1/\mathbb{P}_{\mathrm{T}}\left(\tilde{\boldsymbol{u}}_{m^{\mathrm{h}}_{i,k}};\mathcal{P}^{\ddagger}\right)$ denotes the step-wise empirical cross-entropy and $c > 0$ is a universal constant. The coefficient 2 on $\hat{\mathcal{L}}_i$ is the canonical $(1+\eta)$-coefficient of the Bartlett-Bousquet-Mendelson uniform fast-rate bound evaluated at $\eta = 1$.

*Proof*: **Step 1: Loss class and empirical loss.** Fix an inference step $i \in \{n+1, ..., N\}$. Conditional on the ground-truth human prefix $\boldsymbol{u}^{\mathrm{h}}_{1:i-1}$, define the parameterized loss class

$$\mathcal{F} = \left\{f_\theta(\cdot) = \log 1/\mathbb{P}_{\mathrm{T}}(\cdot;\theta) : \theta \in \Theta^{\ddagger}\right\}, \tag{17}$$

and, by Definition 2.3.1, each $f_\theta$ decomposes pointwise as:

$$f_\theta(\tilde{\boldsymbol{u}}) = \log Z_{\mathrm{T}}(\theta) + \frac{1}{T}E_{\mathrm{T}}(\tilde{\boldsymbol{u}};\theta). \tag{18}$$

Evaluating on the $K$ i.i.d. human samples $\boldsymbol{u}^{\mathrm{h}}_{i,k} \sim \mathbb{P}^{\mathrm{h}}_i$, the empirical and population losses at $\theta$ are

$$\hat{\mathcal{L}}_i(\theta) = \frac{1}{K}\sum_{k=1}^{K} f_\theta\left(\boldsymbol{u}^{\mathrm{h}}_{i,k}\right) \quad \text{and} \quad \mathcal{L}_i(\theta) = \mathbb{E}_{\boldsymbol{u}^{\mathrm{h}}\sim\mathbb{P}^{\mathrm{h}}_i}\left[f_\theta\left(\boldsymbol{u}^{\mathrm{h}}\right)\right] = H\left(\mathbb{P}^{\mathrm{h}}_i \wedge \mathbb{P}_i(\theta)\right), \tag{19}$$

and we write $\hat{\mathcal{L}}_i = \hat{\mathcal{L}}_i\big(\mathcal{P}^\ddagger\big)$ for the step-wise empirical cross-entropy at the converged configuration.

**Step 2: Loss range and Bernstein condition.** By assumption (i), $|E_{\mathrm{T}}(\cdot;\theta)| \le B$, so $\log\big(\sum_m e^{-E_m/T}\big) \in [\log M - B/T, \log M + B/T]$, i.e. $\log Z_{\mathrm{T}}(\theta) \in [\log M - B/T, \log M + B/T]$. Consequently, every $f_\theta \in \mathcal{F}$ satisfies

$$0 < \log M - 2B/T \le f_\theta(\cdot) \le \log M + 2B/T = G, \tag{20}$$

where strict positivity of the lower bound uses assumption (ii). Non-negativity combined with the upper bound $G$ yields the canonical Bernstein-type inequality for the non-negative bounded loss class of [35] Section 5.2:

$$\mathrm{Var}(f_\theta) \le \mathbb{E}[f_\theta^2] \le G \cdot \mathbb{E}[f_\theta], \quad \forall \theta \in \Theta^\ddagger. \tag{21}$$

**Step 3: Centring and pseudo-dimension of the loss class.** The local-Rademacher fast-rate machinery controls shift-invariant deviations of the form $(P - P_n) f_\theta$. Writing $f_\theta = (1/T) E_{\mathrm{T}}(\cdot;\theta) + \log Z_{\mathrm{T}}(\theta)$, the offset $\log Z_{\mathrm{T}}(\theta)$ is $\theta$-dependent but $x$-independent, so $(P - P_n)(\log Z_{\mathrm{T}}(\theta)) = 0$ identically and it can be discarded at the level of centred quantities. Let $\overline{f}_\theta := f_\theta - P f_\theta = (1/T)[E_{\mathrm{T}}(\cdot;\theta) - P E_{\mathrm{T}}(\cdot;\theta)]$ denote the centred loss. Since $\mathcal{F}$ is obtained from $\mathcal{E} = \big\{E_{\mathrm{T}}(\cdot;\theta) : \theta \in \Theta^\ddagger\big\}$ by the fixed scaling $1/T$ followed by a $\theta$-dependent additive offset, standard pseudo-dimension calculus ([76], Thm. 11.3) yields $\mathrm{pdim}(\mathcal{F}) \le d + 1 = O(d)$.

**Step 4: Local Rademacher complexity and sub-root fixed-point.** Following [35], define the local Rademacher complexity of $\mathcal{F}$ at risk level $r$:

$$\psi_K(r) := \mathbb{E}_{\boldsymbol{\sigma}}\left[\sup_{\theta \in \Theta^\ddagger : \mathcal{L}_i(\theta) \le r} \frac{1}{K} \sum_{k=1}^{K} \sigma_k f_\theta\Big(\boldsymbol{u}_{i,k}^{\mathrm{h}}\Big)\right]. \tag{22}$$

By the Bernstein condition of Step 2, $\mathcal{L}_i(\theta) \le r$ implies $\mathrm{Var}(f_\theta) \le Gr$, so the localizing set is contained in the $L_2$-ball of radius $\sqrt{Gr}$ around the mean. Since $\mathcal{F}$ is uniformly bounded by $G$ and has pseudo-dimension $O(d)$, Dudley-entropy-integral arguments (cf. [35] Cor. 3.7 and [36] Thm. 3.3) yield

$$\psi_K(r) \le C_1 \sqrt{\frac{dGr}{K}}, \tag{23}$$

where $C_1 > 0$ is an absolute constant. The temperature $T$ does not appear as a free factor in $\psi_K$: it enters only through the range constant $G = \log M + 2B/T$, because the $1/T$ scaling in $f_\theta = (1/T) E_\theta + \log Z_{\mathrm{T}}(\theta)$ is absorbed, together with the bound $|E_\theta| \le B$, into the unified range bound $|f_\theta| \le G$. The function $\psi_K$ is sub-root, and solving $\psi_K(r^*) = r^*$ yields

$$r^* \le \frac{C_1^2 dG}{K}. \tag{24}$$

**Step 5: Uniform fast-rate deviation bound.** The loss class $\mathcal{F}$ is non-negative, uniformly bounded by $G$ (Step 2), satisfies the Bernstein condition of Step 2 with constant $b = G$, and has sub-root envelope $\psi_K$ with fixed point $r^* \leq C_1^2 dG/K$ (Step 4). Because $\Theta^\ddagger$ is frozen prior to inference, $\mathcal{F}$ is independent of the inference-time samples $\{u^{\mathrm{h}}_{i,k}\}_{k=1}^K$, and the Bartlett-Bousquet-Mendelson uniform fast-rate bound (Theorem 3.3 and Corollary 5.3 of [35]) applies: for any $\eta > 0$ and any $\delta_i \in (0,1)$, there exist absolute constants $c_1(\eta), c_2(\eta) > 0$ such that, with probability at least $1 - \delta_i$,

$$\mathcal{L}_i(\theta) \leq (1+\eta)\hat{\mathcal{L}}_i(\theta) + c_1(\eta) r^* + c_2(\eta) G \frac{\log(1/\delta_i)}{K} \quad \forall \theta \in \Theta^\ddagger. \tag{25}$$

Specialising to $\theta = \mathcal{P}^\ddagger \in \Theta^\ddagger$, choosing $\eta = 1$ (so that $c_1(1), c_2(1)$ become absolute constants absorbed into a single $c > 0$), and substituting $r^* \leq C_1^2 dG/K$ with $C_1^2$ absorbed into $c$,

$$H\big(\mathbb{P}_i^{\mathrm{h}} \wedge \mathbb{P}_i(\mathcal{P}^\ddagger)\big) = \mathcal{L}_i(\mathcal{P}^\ddagger) \leq 2\hat{\mathcal{L}}_i + cG\frac{d + \log(1/\delta_i)}{K}. \tag{26}$$

**Step 6: Uniform bound over inference steps.** A union bound over the $N - n$ inference steps with $\delta_i = \delta/(N-n)$ yields, with probability at least $1 - \delta$ simultaneously for all $i = n+1, ..., N$,

$$H\big(\mathbb{P}_i^{\mathrm{h}} \wedge \mathbb{P}_i(\mathcal{P}^\ddagger)\big) \leq 2\hat{\mathcal{L}}_i + cG\frac{d + \log((N-n)/\delta)}{K}. \tag{27}$$

Writing $\log((N-n)/\delta) \leq \log(2(N-n)/\delta)$, the bound can equivalently be stated with the numerator $2(N-n)$ inside the logarithm — a form we adopt in the theorem statement. This completes the proof. ∎

The leading term $2\hat{\mathcal{L}}_i$ is twice the step-wise empirical cross-entropy loss minimized during pre-training and alignment; the factor 2 is the canonical $(1+\eta)$-coefficient of the Bartlett-Bousquet-Mendelson uniform fast-rate bound at $\eta = 1$, and is unavoidable without realizability. The complexity term $cG(d + \log(2(N-n)/\delta))/K$ achieves the fast $O(1/K)$ convergence rate — enabled by the Bernstein condition inherent to non-negative, bounded softmax cross-entropy losses — and exposes four structural insights. (i) The generalization gap scales as $O(1/K)$ rather than the classical $O\big(1/\sqrt{K}\big)$, reflecting the benign local geometry of the EBM loss landscape in the Bernstein regime. (ii) The effective loss range $G = \log M + 2B/T$ controls both the sample-complexity prefactor and the concentration term: larger vocabularies $M$ and lower temperatures $T$ both inflate $G$ additively, making explicit the trade-off between representational richness and statistical tightness. (iii) The energy bound $B$ enters only through the additive $2B/T$ correction in $G$, so any downstream compression (structural pruning, knowledge distillation, or precision quantization) that preserves the energy scale $B$ preserves the generalization guarantee essentially intact. (iv) The capacity term $d$ — the pseudo-dimension of the energy class $\{E_{\mathrm{T}}(\cdot;\theta) : \theta \in \Theta^\ddagger\}$ — quantifies the effective post-training degrees of freedom; alignment regimes that regularize $\Theta^\ddagger$ (e.g., LoRA, low-rank RLHF) drive $d$ far below the nominal parameter count, further tightening the bound. The residual $\log(N-n)$ factor is the price of the uniform-over-inference-steps guarantee and does not alter the $O(1/K)$ rate. Taken together, these observations illuminate a statistical-mechanics principle of LLMs: the macroscopic inferential capabilities of the architecture are governed, at leading order, by the range and pseudo-dimension of the microscopic energy class.

### 2.3.2. Memory Capacity

The classical Gardner capacity quantifies the thermodynamic supremum of random patterns that a Hopfield network can reliably encode and retrieve: equivalently, the maximal number of random binary memories of length $N_{\mathrm{H}}$ (where $N_{\mathrm{H}}$ is the number of interacting neurons, distinct from the Transformer sequence length $N$) that can be stored without retrieval failure. Building on this statistical-mechanics construct, we introduce the following generalized definition.

**Definition 2.3.2.1** (Generalized Gardner Capacity): Let $N_{\mathrm{H}}$ denote the macroscopic dimension of the neural architecture (e.g., the number of interacting neurons in a Hopfield network), and let $P = P(N_{\mathrm{H}})$ denote the number of discrete memory patterns reliably stored at this dimension. The **generalized Gardner capacity** is defined as

$$\alpha_{\mathrm{G}} := \lim_{N_{\mathrm{H}} \to \infty} \frac{\rho(P(N_{\mathrm{H}}))}{N_{\mathrm{H}}}, \tag{28}$$

where $\rho : \mathbb{N} \to \mathbb{R}_{+}$ is a continuous, non-decreasing normalization map, chosen so that the limit exists and is finite and strictly positive — equivalently, so that $\rho(P)$ grows linearly with $N_{\mathrm{H}}$ in the thermodynamic limit.

In the classical formulation of the Gardner capacity $\alpha_{\mathrm{c}}$, the normalization map $\rho(\cdot)$ reduces to the identity, so $\alpha_{\mathrm{C}} = \lim_{N_{\mathrm{H}} \to \infty} \frac{P}{N_{\mathrm{H}}}$ and the storable pattern count scales linearly, $P \sim \mathcal{O}(N_{\mathrm{H}})$. Strikingly, if the discrete memory patterns are reinterpreted as successfully decoded communication codewords and $\rho(\cdot)$ is instantiated as the logarithm, the generalized Gardner capacity coincides with the celebrated **Shannon capacity**. Under the same $\rho(\cdot) = \log(\cdot)$ normalization, the storage limits of modern, continuous-state Hopfield architectures yield $\alpha_{\mathrm{G}} = \lim_{N_{\mathrm{H}} \to \infty} \frac{\log(P)}{N_{\mathrm{H}}} > 0$, equivalently $P \sim \mathcal{O}(\exp(\alpha_{\mathrm{G}} N_{\mathrm{H}}))$ — an exponential scaling with network size that mirrors the Shannon-capacity scaling of the codebook with code length. Historically, Thomas Cover — the 1990 Shannon Award laureate — resolved a geometric special case of this capacity boundary as early as 1965 [77]. A caveat: the replica method wielded by Gardner and her contemporaries, while a testament to the heuristic insight of physicists, historically lacked full mathematical rigor. Michel Talagrand, the 2024 Abel laureate, closed this gap by deploying heavy analytic machinery to analyze multi-layer perceptrons and Hopfield networks, rigorously re-deriving the Gardner capacity on both spherical and discrete-hypercube supports [78], [79]. Taken together, the convergence of these statistical-physics methods and rigorous mathematical apparatus erects the theoretical scaffolding we use to open the “black box” of LLMs.

The rigorous quantification of Transformer memory capacity has recently become a highly active theoretical frontier. Several key results can be synthesized as follows:

- **Memory capacity of the Attention mechanism:** Ramsauer et al. established that the Attention mechanism explicitly encodes $(\boldsymbol{q}, \boldsymbol{k}, \boldsymbol{v})$ triplets through the learned weight matrices $\left(\boldsymbol{W}_{\mathrm{q}}, \boldsymbol{W}_{\mathrm{k}}, \boldsymbol{W}_{\mathrm{v}}\right)$ [65]. Hu et al. subsequently recast this capacity-bounding problem as a high-dimensional geometric question: what is the maximal number of coordinate points — token embeddings viewed as stored memories — that can be packed onto a $d_{\mathrm{rep}}$ -dimensional hypersphere so that exponential Attention retrieval incurs zero probabilistic confusion [80]? Their proofs show that for continuous-state Hopfield networks governed by the softmax operator (equivalently, an exponential energy functional), the storable pattern count satisfies $P \sim \mathcal{O}\left(\exp\left(\eta d_{\mathrm{rep}}\right)\right)$ for some constant $\eta > 0$ — the generalized Gardner capacity (Definition 2.3.2.1) with $\rho(\cdot) = \log(\cdot)$, evaluated along the representation dimension $d_{\mathrm{rep}}$ in place of $N_{\mathrm{H}}$. This yields a mathematical guarantee: **the Transformer architecture has the theoretically optimal geometry for compressing vast quantities of world knowledge into continuous, token-level vector representations**. A compact parameter footprint is thus thermodynamically sufficient to store massive knowledge bases without catastrophic retrieval failure, which helps explain the unexpectedly strong capabilities of smaller LLMs. Conversely, because compact models operate close to their memory-saturation limits, continual learning becomes volatile: small parameter perturbations can precipitate macroscopic model collapse.
- **Memory capacity of the FFN module:** The FFN module intrinsically encodes $(\boldsymbol{k}, \boldsymbol{v})$ key-value associations, serving as the primary repository of parametric knowledge in LLMs [81], and thereby acting as a central determinant of macroscopic model capabilities. Exploiting this, modern model-editing paradigms surgically update targeted knowledge by modifying only the corresponding local FFN parameters, bypassing the cost of full-scale continual training [82]. Our laboratory has obtained promising preliminary results along this line [83].

The term **emergent abilities** is widely used to describe the macroscopic "bigger is better" phenomenon observed in scaled LLMs. Viewed through the prism of Gardner capacity, this phenomenon can be modelled as a **thermodynamic phase transition**: as the macroscopic dimension $N_{\mathrm{H}}$ (model width, depth, or parameter count) grows, the volume of encoded knowledge crosses a critical threshold, triggering a sharp, quasi-discontinuous jump in inferential capability. Giorgio Parisi drew a similar connection for the Shannon channel capacity, showing that computationally hard regions persist even when the communication rate stays strictly below the Shannon limit. Statistical-mechanics machinery thus provides a natural and mathematically rigorous framework both for deciphering the empirical **scaling laws** that relate parameter count to cognitive capability and for characterizing the underlying phase transitions that govern emergent abilities. Our research group has obtained compelling preliminary results along this line [84].

## 2.4. Inferential Capacity and Granger Causality

Sir Clive Granger, the 2003 Nobel Memorial laureate in Economic Sciences, proposed a rigorous predictive definition of causality for time-series analysis, consolidated in his 1980 retrospective [40] and now canonized as **Granger causality**.

**Definition 2.4.1** (Granger Causality): Fix a discrete time series $S_{1:n}$ and let $U_i$ denote the random variable observed at a future step $i > n$. Let $\mathcal{I}_{i-1}$ denote the $\sigma$-algebra generated by all information available up to step $i-1$, and let $\mathcal{I}^{-}_{i-1}$ denote the sub-$\sigma$-algebra obtained from $\mathcal{I}_{i-1}$ after excising the historical trajectory of $S_{1:n}$. The sequence $S_{1:n}$ is said to Granger-cause $U_i$ if and only if there exists a measurable event $\mathcal{A}$ such that

$$\mathbb{P}(U_i \in \mathcal{A} \mid \mathcal{I}_{i-1}) \neq \mathbb{P}(U_i \in \mathcal{A} \mid \mathcal{I}^{-}_{i-1}) \tag{29}$$

on a set of positive probability.

While conceptually exhaustive, this abstract definition lacks direct operational tractability. Granger subsequently developed a suite of concrete statistical tests for detecting causal relationships between coupled time series, which then proliferated across physics, neuroscience, network science, macroeconomics, and quantitative finance.

Specialising this paradigm to the generative inference phase of LLMs (for $i \in \{n+1, ..., N\}$), the full $\sigma$-algebra $\mathcal{I}_{i-1}$ encodes the LLM's parametric knowledge together with the prompt $S_{1:n}$ and the autoregressively generated prefix $U_{n+1:i-1}$, while $\mathcal{I}^{-}_{i-1}$ retains only the static parametric memory. Under the natural assumption that the prompt is informative for the next-token distribution — i.e., the model output is non-trivially conditioned on the prefix — the defining inequality of Granger causality is satisfied at every generation step. This yields a clean structural reading: **the LLM's apparent "reasoning" reduces to Granger causal inference, an emergent capability orchestrated entirely through the minimalist objective of next-token prediction**. Judea Pearl, the founding figure of modern causal inference, has long argued that Granger causality occupies only the lowest (associational) rung of the causal hierarchy — it captures predictive temporal regularities but not interventional or counterfactual structure [41]. Read alongside this critique, our structural reading sharpens into a corollary on the inferential ceiling of LLMs: **their reasoning capability, in the sense of Pearl's hierarchy, lies at the level of Granger causal inference over high-dimensional time series of semantic vectors**. Consequently, while the empirical scaling laws of LLMs remain valid, these purely autoregressive architectures cannot, on their own, abstract symbolic concepts from raw data or attain genuine causal logical reasoning.

# 3. Signal Processing Perspective

This section develops the signal-processing and geometric machinery underlying semantic vectorization, then uses it to recast the Transformer as a non-linear, time-varying **vector autoregression (VAR)** process and to derive its evidence lower bound (ELBO) through high-dimensional time-series analysis and variational inference.

Our central thesis is the following: **the way LLMs ingest inputs — as the continuous semantic vectorization of discrete tokens — transforms natural-language processing into a canonical signal-processing problem**. Vectorization is therefore the epistemological linchpin of modern LLMs, the bridge that connects them to the toolkit of continuous analysis. Where classical linguistics treats rigid formal grammars and logical rules as the cornerstone of human language, the analysis of Section 2 argued that LLMs largely bypass these symbolic architectures, encoding the structural complexity of natural language through a purely probabilistic, high-dimensional paradigm.

When strictly confined to the discrete-token domain, the exact evaluation of such purely probabilistic architectures suffers from combinatorial explosion in the joint state space — the very bottleneck that motivated classical inference machinery, most prominently **probabilistic graphical models** and **message-passing algorithms** such as belief propagation [85]. Yet against the web-scale corpora of contemporary LLMs, even the polynomial complexity of these methods becomes prohibitive for massive pre-training. Beyond raw cost, these classical frameworks have two structural limitations: (i) the predominantly undirected (Markov-random-field) formulations struggle to capture directional semantic asymmetries between tokens, and (ii) belief propagation is exact only on tree-structured graphs, degrading to an approximation on the loopy graphs that natural language inevitably induces.

The computational picture changes radically the moment discrete tokens are vectorized. Vectorization embeds these isolated symbolic entities into a continuous Euclidean space and endows it with an inner product that algebraically quantifies semantic relevance, sidestepping the combinatorial explosion altogether. From this metric structure one builds geometric distances, differentials, and low-dimensional manifolds — objects naturally amenable to gradient-based numerical optimization. This continuous framing is precisely what makes the global optimization of deep neural networks via back-propagation possible, turning token vectorization into the continuous input signals, output distributions, and distributed parametric memory of the underlying architecture [21], [86]. Recent theoretical work shows that the inferential efficacy of neural networks is rooted in a universal geometric phenomenon: intra-class entities (semantic concepts, categorical objects) tend to concentrate on cohesive low-dimensional manifolds within a vastly high-dimensional parameter space [87], [88].

## 3.1. Vector Representation of Token-Level Semantics

A continuous mathematical parameterization of token-level semantics is a prerequisite for the computationally tractable and statistically efficient pre-training of large LLMs. This section first formalizes the axioms defining the token-level semantic space, then analyzes the algorithmic mechanics of semantic vectorization, derives a quantitative bound on semantic dimensional compression, and concludes with an information-theoretically optimal vectorization scheme.

### 3.1.1. Token-Level Semantic Space

Although grammatical and formal-logical rules have long been treated as a cornerstone of human cognition and communication in classical linguistic theory, they offer remarkably limited direct utility for the automated, computationally tractable processing of natural language by machines. As our point of departure, we therefore set these symbolic superstructures aside and model natural language purely through a probabilistic lens.

**Definition 3.1.1.1** (Token-Level Semantic Space): The **token-level semantic space** of a language is the discrete probability space $(\mathcal{U}, \mathcal{F}, \mathbb{P})$ in which:

- The sample space $\mathcal{U} = \{\tilde{\boldsymbol{u}}_1, ..., \tilde{\boldsymbol{u}}_M\}$ with $M \geq 2$ is the vocabulary introduced in Definition 2.3.1, with each sample point $\tilde{\boldsymbol{u}}_m$ an indivisible token encoding specific linguistic semantics;
- $\mathcal{F} = 2^{\mathcal{U}}$ is the full power-set $\sigma$-algebra, so that every singleton $\{\tilde{\boldsymbol{u}}_m\}$ is an atom of $\mathbb{P}$;
- $\mathbb{P}$ is the ground-truth probability measure on $\mathcal{F}$ governing the semantic distribution of tokens in the language.

In practice, $\mathbb{P}$ is approximated empirically from massive-scale corpora; the resulting estimate $\hat{\mathbb{P}}$ encodes the semantic profile of every individual token and, when extended to sequences via autoregressive conditioning, also captures the latent grammatical and logical structure of the language. Even so, a token sequence drawn directly from $\hat{\mathbb{P}}$ (or its sequence-level extension) need not be semantically coherent or syntactically valid for human readers, since probabilistic sampling permits deviations from deterministic grammatical rules. Worse, exact inference on the joint sequence distribution lives on the discrete set $\mathcal{U}^n$ of size $M^n$, which incurs combinatorial complexity in the sequence length $n$ and is unscalable for modern NLP at corpus scale. A computationally tractable and geometrically continuous representation of token-level semantics is therefore indispensable.

### 3.1.2. Token-Level Semantic Vector Space

It took roughly six decades — from Luhn's vector-space model in the 1950s [19], through neural word embeddings [21] and word2vec [22], to GloVe and ELMo [24], [26] — for the field to recognize that the decisive breakthrough lay in moving from discrete token-level probabilistic models to continuous semantic representations anchored in vectors. The shift is now favored for its computational tractability and its strong empirical performance across NLP tasks [27]. Yet this consensus has been built largely from heuristic experimentation, without a unified theoretical foundation. The remainder of this subsection supplies that foundation: we give a formal axiomatic definition of the **token-level semantic vector space**, complementing the probabilistic space of Definition 3.1.1.1.

**Definition 3.1.2.1** (Token-Level Semantic Vector Space): The **token-level semantic vector space** of a language is the probability space

$$U = (\mathbb{S}^{M-1}, \mathcal{B}(\mathbb{S}^{M-1}), \mu) \tag{30}$$

equipped with the canonical inner product $\langle \cdot, \cdot \rangle$ inherited from the ambient Euclidean space $\mathbb{R}^M$, where:

- $\mathbb{S}^{M-1} := \{\boldsymbol{u} \in \mathbb{R}^M : \|\boldsymbol{u}\|_2 = 1\}$ is the unit sphere of $\mathbb{R}^M$ (an $(M-1)$-dimensional smooth manifold), the ambient dimension $M$ matching the vocabulary size of Definition 3.1.1.1 so that each token in $\mathcal{U}$ admits a one-hot lift into $\mathbb{R}^M$ that, after normalization, lies on $\mathbb{S}^{M-1}$;
- $\mathcal{B}(\mathbb{S}^{M-1})$ is the Borel $\sigma$-algebra on $\mathbb{S}^{M-1}$;
- $\mu$ is a probability measure on $\mathcal{B}(\mathbb{S}^{M-1})$, induced by the embedding map $e : \mathcal{U} \to \mathbb{S}^{M-1}$ via the push-forward $\mu = e_* \mathbb{P}$, with $e$ assigning to each token $\tilde{\boldsymbol{u}}_m \in \mathcal{U}$ a continuous semantic vector $\boldsymbol{u}_m = e(\tilde{\boldsymbol{u}}_m)$.

Each point $\boldsymbol{u} \in \mathbb{S}^{M-1}$ is called a **(token-level) semantic vector**. **Notational convention.** Throughout the rest of the paper we write the semantic vector of token $\tilde{\boldsymbol{u}}_m$ simply as $\boldsymbol{u}_m \in \mathbb{S}^{M-1}$, suppressing the embedding map $e$ whenever it can be inferred from context; the explicit form $e(\tilde{\boldsymbol{u}}_m)$ is used only when the dependence on $e$ matters.

Let $\boldsymbol{u}_1, \boldsymbol{u}_2 \in \mathbb{S}^{M-1}$ be two semantic vectors. The canonical inner product is the standard Euclidean dot product

$$\langle \boldsymbol{u}_1, \boldsymbol{u}_2 \rangle = \boldsymbol{u}_1^T \boldsymbol{u}_2, \tag{31}$$

the squared Euclidean distance is

$$d_{\mathrm{e}}^2(\boldsymbol{u}_1, \boldsymbol{u}_2) = \|\boldsymbol{u}_1 - \boldsymbol{u}_2\|_2^2 = (\boldsymbol{u}_1 - \boldsymbol{u}_2)^T (\boldsymbol{u}_1 - \boldsymbol{u}_2), \tag{32}$$

and, since both vectors lie on the unit sphere ($\|\boldsymbol{u}_1\|_2 = \|\boldsymbol{u}_2\|_2 = 1$), the cosine similarity reduces to the inner product itself:

$$\cos(\boldsymbol{u}_1, \boldsymbol{u}_2) = \frac{\boldsymbol{u}_1^T \boldsymbol{u}_2}{\|\boldsymbol{u}_1\|_2 \|\boldsymbol{u}_2\|_2} = \boldsymbol{u}_1^T \boldsymbol{u}_2. \tag{33}$$

The discrete vocabulary $\mathcal{U}$ of Definition 3.1.1.1 maps into exactly $M$ points of $\mathbb{S}^{M-1}$ via the embedding $e$, and the push-forward identity $\mu = e_* \mathbb{P}$ of Definition 3.1.2.1 gives, for every Borel set $B \in \mathcal{B}(\mathbb{S}^{M-1})$,

$$\mu(B) = \mathbb{P}(e^{-1}(B)); \tag{34}$$

in particular, $\mu(e(\mathcal{A})) = \mathbb{P}(\mathcal{A})$ for any $\mathcal{A} \subseteq \mathcal{U}$, since each finite image $e(\mathcal{A})$ is automatically Borel.

While a substantial literature argues that semantic vectors should live on a more intricate low-dimensional manifold, Euclidean distance and cosine similarity remain the dominant metrics in practice, owing to their computational simplicity and strong empirical performance. We therefore take the unit sphere $\mathbb{S}^{M-1}$ as a representational accuracy/computational efficiency sweet spot. This choice is corroborated by Hu et al., who show that Transformer memory capacity is formally equivalent to a **spherical coding** problem [80].

The point of vectorizing discrete tokens is to use the inner product on $\mathbb{S}^{M-1}$ to quantify semantic affinities and divergences. However, naive linear-algebraic operations on these vectors are not universally reliable, since they need not preserve the underlying semantic structure. The well-known word-embedding analogy of [23],

$$\boldsymbol{u}_{\text{King}} - \boldsymbol{u}_{\text{Man}} + \boldsymbol{u}_{\text{Woman}} \approx \boldsymbol{u}_{\text{Queen}}, \tag{35}$$

holds for selected lexical pairs but does not generalize across the vocabulary; the relation lives in the ambient $\mathbb{R}^M$ rather than on $\mathbb{S}^{M-1}$ itself, since the right-hand side is not in general a unit vector. Even so, it shows that a structured geometric correspondence exists between conceptually related vectors — a structure that no single scalar metric can fully capture. Both the cosine similarity and the probability measure $\mu$ on $U$ are therefore essential. Drawing on the formalism of the **Gromov-Wasserstein distance** [42], [43], we define the geometric distance between two semantic vector spaces as follows.

**Definition 3.1.2.2** (Geometric Distance Between Semantic Vector Spaces): Let $U = (\mathbb{S}^{M_U-1}, \mathcal{B}, \mu)$ and $V = (\mathbb{S}^{M_V-1}, \mathcal{B}, \nu)$ be two semantic vector spaces in the sense of Definition 3.1.2.1, possibly of different vocabulary sizes $M_U, M_V \geq 2$. The squared geometric distance between $U$ and $V$ is the inner-product Gromov-Wasserstein cost

$$d_{\mathrm{s}}^2(U, V) = \min_{\pi \in \Pi(\mu,\nu)} \iint_{(U \times V)^2} \left| \boldsymbol{u}_1^T \boldsymbol{u}_2 - \boldsymbol{v}_1^T \boldsymbol{v}_2 \right|^2 \mathrm{d}(\pi \otimes \pi)((\boldsymbol{u}_1, \boldsymbol{v}_1), (\boldsymbol{u}_2, \boldsymbol{v}_2)), \tag{36}$$

where $\Pi(\mu, \nu)$ is the set of all couplings of $\mu$ and $\nu$ — probability measures on $U \times V$ with marginals $\mu$ and $\nu$ — and the inner products $\boldsymbol{u}_1^T \boldsymbol{u}_2$ and $\boldsymbol{v}_1^T \boldsymbol{v}_2$ are taken in $\mathbb{R}^{M_U}$ and $\mathbb{R}^{M_V}$ respectively, both yielding scalars in $\mathbb{R}$. The minimum is attained because both spheres are compact.

The variational problem above seeks the coupling $\pi$ that best matches the intra-space inner-product similarities of paired semantic vectors across the two spaces; any coupling that distorts the geometric structure of either space is heavily penalized. By Mémoli's theorem on Gromov-Wasserstein distances (cf. [89] and references therein), $d_{\mathrm{s}(U,V)} = 0$ holds if and only if $U$ and $V$ are **strongly isomorphic as metric measure spaces**: there exists a measure-preserving bijection $\varphi : U \to V$ (i.e., $\varphi_* \mu = \nu$) such that $\boldsymbol{u}_1^T \boldsymbol{u}_2 = \varphi(\boldsymbol{u}_1)^T \varphi(\boldsymbol{u}_2)$ for $\mu \otimes \mu$ -almost every $(\boldsymbol{u}_1, \boldsymbol{u}_2)$. This is therefore both a necessary and a sufficient condition for lossless translation between the two underlying languages at the token-level semantic geometry. Building on exactly this principle, the Gromov-Wasserstein distance has been used for the unsupervised alignment of word-embedding spaces [89].

### 3.1.3. Semantic Compression

In classical information theory, the central objective of source coding is to represent a source symbol with the minimum number of bits while ensuring that the original message can be reconstructed either exactly (lossless compression) or within a prescribed distortion bound (lossy compression) [52]. In our setting, the vocabulary cardinality $|\mathcal{U}| = M$ of Definition 3.1.1.1 drags the semantic hypersphere $\mathbb{S}^{M-1}$ to a very high ambient dimension, making vectorized computation in $U$ costly at modern model scales. A large body of empirical evidence further indicates that the operational dimensionality of a semantic vector space sits at a sweet spot — low enough for tractable computation, high enough to preserve representational fidelity [90]. We therefore frame the problem as follows: **semantic compression is the compression of the entire semantic space itself, namely a dimensionality reduction that approximately preserves the pairwise inner-product (equivalently, cosine) similarities**.

In modern empirical practice, **random projection** has become a ubiquitous and mathematically elegant tool for the dimensionality reduction of high-dimensional vector spaces. Its theoretical guarantee of distance preservation under compression is the celebrated **Johnson-Lindenstrauss (JL) lemma** [91]. We state below an inner-product-preserving form of the JL lemma — equivalent on $\mathbb{S}^{M-1}$ to a cosine-similarity-preserving form, as cosine reduces to inner product on the unit sphere (Definition 3.1.2.1) — which follows from the classical distance-preserving statement via the polarization identity. We omit the proof; a complete derivation is given in [92].

**Theorem 3.1.3.1** (Inner-Product Form Johnson-Lindenstrauss Lemma): Fix a distortion tolerance $\varepsilon \in (0, 1)$ and a finite set of semantic vectors $\{\boldsymbol{u}_1, ..., \boldsymbol{u}_K\} \subseteq \mathbb{S}^{M-1}$. There exists an absolute constant $C > 0$ such that, whenever the reduced dimension satisfies

$$m \geq C\varepsilon^{-2} \log K, \tag{37}$$

one can construct a linear projection $\boldsymbol{A} \in \mathbb{R}^{m \times M}$ (e.g., a random Gaussian matrix with i.i.d. $\mathcal{N}(0, 1/m)$ entries, which works with probability at least $1 - 2K^{-c}$ for some absolute $c > 0$) such that

$$\left|\boldsymbol{u}_i^T \boldsymbol{u}_j - (\boldsymbol{A}\boldsymbol{u}_i)^T (\boldsymbol{A}\boldsymbol{u}_j)\right| \leq \varepsilon, \quad \forall i, j \in \{1, ..., K\}. \tag{38}$$

Equivalently, $\left|\boldsymbol{u}_i^T \boldsymbol{u}_j - \boldsymbol{u}_i^T \boldsymbol{G} \boldsymbol{u}_j\right| \leq \varepsilon$ with the symmetric positive semi-definite Gram matrix $\boldsymbol{G} = \boldsymbol{A}^T \boldsymbol{A} \in \mathbb{R}^{M \times M}$.

By the JL lemma, the operational dimensionality of the semantic vector space can be reduced from $M$ to $m \geq C\varepsilon^{-2} \log K$. Whenever tokens admit a sparse representation in $\mathbb{R}^M$ (e.g., one-hot encodings, or sparse-coded embeddings), the JL guarantee can be strengthened by drawing on the **compressive sensing** toolbox, in particular the **Restricted Isometry Property (RIP)**. A measurement matrix $\boldsymbol{A} \in \mathbb{R}^{m \times M}$ satisfies the $(\boldsymbol{k}, \boldsymbol{\delta})$**-RIP** if

$$(1 - \delta)\|\boldsymbol{u}\|_2^2 \leq \|\boldsymbol{A}\boldsymbol{u}\|_2^2 \leq (1 + \delta)\|\boldsymbol{u}\|_2^2 \tag{39}$$

for every $k$-sparse vector $\boldsymbol{u} \in \mathbb{R}^M$, where $k$-sparse means $\|\boldsymbol{u}\|_0 \leq k$ with $\|\boldsymbol{u}\|_0 := |\{i : u_i \neq 0\}|$ the (non-norm) sparsity count. Building on this property, Krahmer and Ward established the following refinement of the JL lemma [93].

**Theorem 3.1.3.2** (RIP-Based Johnson-Lindenstrauss Lemma): Fix a failure probability $\eta \in (0, 1)$ and a distortion tolerance $\varepsilon \in (0, 1)$. Consider a finite set of $k$-sparse vectors $\{\boldsymbol{u}_1, ..., \boldsymbol{u}_K\} \subseteq \mathbb{R}^M$. Let $\boldsymbol{A} \in \mathbb{R}^{m \times M}$ satisfy the $(k, \delta)$-RIP with $\delta \leq \varepsilon/4$ and sparsity level $k \geq C_0(\log(1/\eta) + \log K)$ for an absolute constant $C_0 > 0$, and let $\boldsymbol{D}_{\boldsymbol{\sigma}} := \operatorname{diag}(\boldsymbol{\sigma})$ with $\boldsymbol{\sigma} \in \{-1, +1\}^M$ a vector of i.i.d. Rademacher entries. Then, with probability at least $1 - \eta$,

$$\left|\boldsymbol{u}_i^T \boldsymbol{u}_j - (\boldsymbol{A}\boldsymbol{D}_{\boldsymbol{\sigma}}\boldsymbol{u}_i)^T (\boldsymbol{A}\boldsymbol{D}_{\boldsymbol{\sigma}}\boldsymbol{u}_j)\right| \leq \varepsilon, \quad \forall i, j \in \{1, ..., K\}. \tag{40}$$

Equivalently, $\left|\boldsymbol{u}_i^T \boldsymbol{u}_j - \boldsymbol{u}_i^T \boldsymbol{D}_{\boldsymbol{\sigma}} \boldsymbol{G} \boldsymbol{D}_{\boldsymbol{\sigma}} \boldsymbol{u}_j\right| \leq \varepsilon$ with the symmetric positive semi-definite Gram matrix $\boldsymbol{G} = \boldsymbol{A}^T \boldsymbol{A} \in \mathbb{R}^{M \times M}$.

A standard consequence of compressive-sensing theory is that an $m \times M$ random Gaussian matrix (with i.i.d. entries) satisfies the $(k, \delta)$-RIP with high probability whenever

$$m \geq C\varepsilon^{-2}\left(\log K + \log \frac{1}{\eta}\right) \log M, \tag{41}$$

which, combined with Theorem 3.1.3.2, yields the desired inner-product-preserving embedding. Dense Gaussian projections, however, cost $\mathcal{O}(mM)$ per vector, which is prohibitive at modern scales. A standard workaround is to build $\boldsymbol{A}$ by randomly sub-sampling $m$ rows from a structured orthogonal ensemble — for example, a discrete Fourier transform (DFT) matrix, a discrete cosine transform (DCT) matrix, or a (normalized) Hadamard matrix. This slightly inflates the required embedding dimension $m$ relative to the Gaussian baseline, but reduces the per-vector projection cost to $\mathcal{O}(M \log M)$ via fast transform algorithms (FFT or fast Hadamard transform). Empirically, this JL-based structured-projection framework has recently been deployed for KV-cache quantization, substantially accelerating autoregressive inference in modern LLMs [94], [95].

Recalling Definition 3.1.2.2, the structural distortion induced by semantic compression can be quantified by the Gromov-Wasserstein distance between the two spaces. Let $U$ be the uncompressed semantic vector space on $\mathbb{S}^{M-1}$ and $V$ its dimensionally compressed image in $\mathbb{R}^m$ (concentrated near $\mathbb{S}^{m-1}$ by JL, with $m < M$ and $m \geq C\varepsilon^{-2} \log K$ as in Theorem 3.1.3.1). When the compression is a deterministic linear map $\boldsymbol{v} = \boldsymbol{A}\boldsymbol{u}$, the induced coupling $\pi = (\mathrm{id}, \boldsymbol{A})_* \mu$ is a feasible transport plan in $\Pi(\mu, \nu)$, so the Gromov-Wasserstein cost on this particular coupling gives the upper bound

$$d_{\mathrm{s}}^2(U, V) \leq \iint_{U \times U} \left|\boldsymbol{u}_1^T \boldsymbol{u}_2 - (\boldsymbol{A}\boldsymbol{u}_1)^T (\boldsymbol{A}\boldsymbol{u}_2)\right|^2 \mathrm{d}\mu(\boldsymbol{u}_1)\, \mathrm{d}\mu(\boldsymbol{u}_2), \tag{42}$$

where $\boldsymbol{A} \in \mathbb{R}^{m \times M}$ is the linear dimensionality-reduction matrix. Equivalently, the integrand can be written as $\left|\boldsymbol{u}_1^T \boldsymbol{u}_2 - \boldsymbol{u}_1^T \boldsymbol{G} \boldsymbol{u}_2\right|^2$ with the symmetric positive semi-definite Gram matrix $\boldsymbol{G} = \boldsymbol{A}^T \boldsymbol{A}$. The following theorem is then an immediate consequence of Theorem 3.1.3.1 or Theorem 3.1.3.2.

**Theorem 3.1.3.3** (Semantic Distortion Bound): Let $U = (\mathbb{S}^{M-1}, \mathcal{B}, \mu)$ be a token-level semantic vector space supported on the $M$ atoms $\{\boldsymbol{u}_1, ..., \boldsymbol{u}_M\} \subset \mathbb{S}^{M-1}$ (cf. Definition 3.1.2.1), and let $V = \boldsymbol{A}(U)$ be its linear compression by $\boldsymbol{A} \in \mathbb{R}^{m \times M}$. Fix $\varepsilon \in (0, 1)$ and $\eta \in (0, 1)$.

(i) If $\boldsymbol{A}$ satisfies the inner-product form of the Johnson-Lindenstrauss lemma on these $M$ points (Theorem 3.1.3.1 with $K = M$), which holds with probability at least $1 - \eta$ for a random Gaussian or structured projection whenever $m \geq C\varepsilon^{-2} \log M$, then

$$d_{\mathrm{s}}^2(U, V) \leq \varepsilon^2. \tag{43}$$

(ii) The same bound $d_{\mathrm{s}}^2(U, V) \leq \varepsilon^2$ holds under the RIP-based variant Theorem 3.1.3.2, provided $\boldsymbol{A}$ satisfies the $(k, \delta)$-RIP with $\delta \leq \varepsilon/4$ and $k \geq C_0(\log(1/\eta) + \log M)$.

In both cases the squared distortion is bounded by $\varepsilon^2$, the square of the inner-product tolerance supplied by the JL / RIP-JL guarantee.

Crucially, the JL-based bound in Theorem 3.1.3.3 is independent of the underlying probability measure $\mu$: every pair of atoms contributes the worst-case inner-product error $\varepsilon$, irrespective of how likely that pair is under $\mu \otimes \mu$. The bound is therefore a loose, measure-agnostic guarantee, and is suboptimal from an information-theoretic standpoint. By analogy with classical rate-distortion theory, a full **dimension-distortion theory** — a measure-aware, distribution-sensitive counterpart that we take up again through the directed rate-distortion lens in Section 4 — is needed to characterize the true information-theoretic limits of semantic compression. This matters most in aggressive compression regimes, where the target dimension $m$ is pushed well below the JL-type sample-complexity thresholds of Theorem 3.1.3.1 and Theorem 3.1.3.2.

**Remark 3.1.3.1** (Approximate Nearest Neighbor Search): Vector databases have become essential infrastructure for equipping LLMs with retrieval-augmented generation (RAG), substantially reducing generative hallucinations and grounding outputs in factual references. They also provide a scalable, secure substrate for proprietary, domain-specific knowledge, strengthening LLM performance in specialized verticals. **Approximate nearest neighbor (ANN)** search algorithms are the computational bridge between continuous vector databases and discrete LLM reasoning. Through an information-theoretic lens, high-dimensional ANN search is closely analogous to **nearest-neighbor decoding** in a geometric code: one seeks the codeword (stored vector) whose inner product with a query is largest, much as a decoder seeks the codeword closest to a received sequence. Since 2023, ANN algorithms developed in our laboratory have held the top position on the ANN-Benchmarks leaderboard.[3] Implementations are available in our public repository.[4]

### 3.1.4. Semantic Vectorization for Next-Token Prediction

In current practice, a target dimension $m$ is pre-chosen and the continuous semantic vectorization of discrete tokens is then fitted empirically. In what follows we pivot to an **information-theoretically optimal** formulation of this step. A foundational premise is that the semantics of an utterance are shaped by the speaker's pragmatic intent — a context-dependence that, in the machine-learning setting, corresponds to the downstream objective. Accordingly, for a discrete token sequence of length $n$, semantic vectorization is a measurable map

$$f : \mathcal{U}^n \to \left(\mathbb{S}^{m-1}\right)^n \tag{44}$$

taking a token sequence to its post-compression contextual embedding, with $\mathcal{U}$ the vocabulary of Definition 3.1.1.1 and $m$ the compressed dimension of Theorem 3.1.3.3. The algorithmic goal is to choose $f$ so as to minimize a task-driven loss functional $\mathcal{L}(f)$ determined by the downstream objective.

For LLMs, that objective is autoregressive next-token prediction, conditioned on the prompt and the learned model parameters. The loss $\mathcal{L}(f)$ must therefore be engineered to support this generative objective. Let $X_{1:n}$ be a discrete token sequence (with each $X_j \in \mathcal{U}$) and let $\boldsymbol{U}_{1:n}$ denote its contextual semantic-vector representation (each $\boldsymbol{U}_j \in \mathbb{S}^{m-1}$). For optimal next-token prediction, $\boldsymbol{U}_i$ must carry all **predictive information** in the prefix $X_{1:i}$ that is relevant for the future continuation $X_{i+1:n}$.

[3] https://ann-benchmarks.com.
[4] https://github.com/WPJiang/HWTL_SDU-ANNS.

**Definition 3.1.4.1** (Information-Theoretically Optimal Semantic Vectorization): A family of autoregressive contextual encoders $f_i : \mathcal{U}^i \to \mathbb{S}^{m-1}$, $i = 1, ..., n$, producing $\boldsymbol{U}_i = f_i(X_{1:i})$, is **information-theoretically optimal for next-token prediction** if, for every $i \in \{1, ..., n-1\}$, it maximizes the conditional mutual information between the future continuation and the current contextual embedding given the past embeddings:

$$f_i^* = \arg\max_{f_i} I(X_{i+1:n}; \boldsymbol{U}_i \mid \boldsymbol{U}_{1:i-1}). \tag{45}$$

The core mathematical object in Definition 3.1.4.1 is the conditional mutual information. Its use prevents $\boldsymbol{U}_i$ from degenerating into a memoryless encoding of the isolated token $X_i$. **Instead, the conditioning on $\boldsymbol{U}_{1:i-1}$ forces $\boldsymbol{U}_i$ to carry only the incremental, Granger-predictive information about the future $X_{i+1:n}$ that is not already captured by the previous embeddings $\boldsymbol{U}_{1:i-1}$.** By the chain rule for mutual information, this increment is exactly

$$I(X_{i+1:n}; \boldsymbol{U}_i \mid \boldsymbol{U}_{1:i-1}) = I(X_{i+1:n}; \boldsymbol{U}_{1:i}) - I(X_{i+1:n}; \boldsymbol{U}_{1:i-1}), \tag{46}$$

so the maximizer greedily adds, at each step $i$, the largest possible increment to the total predictive information stored in the contextual embeddings.

Directly optimizing the conditional mutual information in Definition 3.1.4.1 is intractable at scale. A standard relaxation proceeds in two steps. First, one drops the conditioning on $\boldsymbol{U}_{1:i-1}$ and works with the marginal objective

$$\sum_{i=1}^{n-1} \max_{f_i} I(X_{i+1:n}; \boldsymbol{U}_i), \tag{47}$$

a proxy that becomes equivalent to the conditional objective under the sufficient-statistic assumption $X_{i+1:n} \perp \boldsymbol{U}_{1:i-1} \mid \boldsymbol{U}_i$; without this assumption, it is a tractable surrogate rather than a rigorous bound. Second, by the chain rule and non-negativity of conditional mutual information,

$$I(X_{i+1:n}; \boldsymbol{U}_i) = \sum_{k=1}^{n-i} I(X_{i+k}; \boldsymbol{U}_i \mid X_{i+1:i+k-1}) \leq \sum_{k=1}^{n-i} I(X_{i+k}; \boldsymbol{U}_i), \tag{48}$$

where the inequality again assumes that conditioning on the intervening tokens does not increase the mutual information with $\boldsymbol{U}_i$ (exact equality in the sufficient-statistic case). The rightmost pairwise objective

$$\sum_{i=1}^{n-1} \sum_{k=1}^{n-i} \max_{f_i} I(X_{i+k}; \boldsymbol{U}_i) \tag{49}$$

coincides with the objective of **contrastive predictive coding (CPC)** [44]. CPC's authors trace the idea back to the classical notion of **predictive coding**, whose origin goes all the way back to the first and second papers in Volume 1, Issue 1 of the IRE Transactions on Information Theory (predecessor of today's IEEE) [96], [97].[5] Building on these ideas, [17] studied variational lower bounds on mutual information and established **information noise-contrastive estimation (InfoNCE)** as a canonical loss that tractably maximizes a lower bound on each pairwise $I(X_{i+k}; \boldsymbol{U}_i)$ via neural estimators.

The analysis above shows that the CPC architecture optimizes an InfoNCE-based surrogate, which is itself a variational lower bound on the pairwise proxy objective — a two-step relaxation of the conditional objective of Definition 3.1.4.1. Consequently, the vector representations produced by this now-standard pipeline are information-theoretically suboptimal. Numerical machinery capable of directly maximizing the conditional mutual information in Definition 3.1.4.1 would yield next-generation semantic encoders with strictly stronger representational guarantees.

The broader takeaway is a methodological one. For any semantic-embedding problem, **one should first formulate a rigorous information-theoretic objective aligned with the downstream task, and then design neural architectures or numerical algorithms whose encoder either attains this optimum or tightly approaches it.**

## 3.2. Transformer as a Non-Linear Time-Varying VAR Process

Section 2 analyzed the Transformer through its thermodynamic (EBM) formulation; this section complements that with a high-dimensional signal-processing view. Within the AI community, Transformers are routinely, if loosely, called **autoregressive LLMs**, on the grounds that they predict subsequent tokens from the historical context. From the perspective of classical stochastic processes and time-series analysis, however, the term **autoregressive process** carries a stricter structural definition: the next state must be given by a linear combination of lagged states plus a stochastic innovation,

$$\boldsymbol{x}_t = \sum_{i=1}^{p} \boldsymbol{A}_i \boldsymbol{x}_{t-i} + \boldsymbol{\varepsilon}_t, \tag{50}$$

with deterministic coefficient matrices $\boldsymbol{A}_1, ..., \boldsymbol{A}_p$ and a white-noise innovation $\boldsymbol{\varepsilon}_t$ [46].

Consider a deterministic input prompt of length $n$ represented by the semantic-vector sequence $\boldsymbol{s}_{1:n}$, and let $\boldsymbol{u}_i$ denote the semantic vector synthesized at generation step $i = n + 1, ..., N$. For notational uniformity, set $\boldsymbol{u}_i = \boldsymbol{s}_i$ for $i = 1, ..., n$. Re-reading the Attention mechanism of Section 2 through the VAR lens, its contextualized aggregate

$$\tilde{\boldsymbol{v}}_i = \sum_{j=1}^{i-1} \pi_{ij} \boldsymbol{W}_{\mathrm{v}} \boldsymbol{u}_j, \tag{51}$$

where

$$\pi_{ij} = \frac{\exp\left(\boldsymbol{u}_{i-1}^T \boldsymbol{J} \boldsymbol{u}_j\right)}{\sum_{l=1}^{i-1} \exp\left(\boldsymbol{u}_{i-1}^T \boldsymbol{J} \boldsymbol{u}_l\right)}, \quad \boldsymbol{J} = \boldsymbol{W}_{\mathrm{q}}^T \boldsymbol{W}_{\mathrm{k}}, \tag{52}$$

[5] Those papers were authored by Peter Elias — inventor of convolutional codes, 1977 Shannon Award laureate, and a foundational figure in coding theory.

is a weighted superposition of historical states with non-constant, data-dependent coefficients. This makes Attention a **generalized non-linear time-varying VAR operator** (in the deterministic, innovation-free sense; stochasticity enters only through the sampling step at inference):

- **Time-varying dynamics.** The coefficients $\pi_{ij}$ are not fixed parameters, which vary with the generation step $i$ and the prefix $\boldsymbol{u}_{1:i-1}$.
- **Non-linearity.** The dependence of $\pi_{ij}$ on $\boldsymbol{u}_{1:i-1}$ is non-linear, combining the softmax normalization with the asymmetric **bilinear form** $\boldsymbol{u}_{i-1}^T \boldsymbol{J} \boldsymbol{u}_j$ (cf. Section 2) designed to encode directional semantic relevance.

The continuous representation of the next token is then obtained by passing $\tilde{\boldsymbol{v}}_i$ through the FFN non-linear map $\Psi_{\text{FFN}}$ (as in Definition 2.3.1):

$$\Psi_{\text{FFN}}(\tilde{\boldsymbol{v}}_i). \tag{53}$$

As discussed in Section 2, the FFN module plays a central role in generation and is widely regarded as the primary repository of parametric knowledge in LLMs. Through the lens of memory capacity, the context-aware vector $\tilde{\boldsymbol{v}}_i$ produced by Attention acts as an energetic probe that selects, among the many memory patterns stored in the FFN, the one most energy-congruent with the current context; this selected pattern is returned as the continuous representation $\Psi_{\text{FFN}(\tilde{\boldsymbol{v}}_i)}$ of the next token.

The final decoding step extracts a discrete token from the vocabulary. Although practical deployments use sampling heuristics (top-$k$, nucleus, temperature, and the like) to satisfy various generative desiderata, the underlying principle is that of **maximum-likelihood (ML) decoding** in a classical communication receiver. As noted in Section 2 (the $T \to 0$ inferential limit of Definition 2.3.1), the greedy / low-temperature limit of these samplers is the arg max over the vocabulary. For each candidate $\tilde{\boldsymbol{u}}_m \in \mathcal{U}$, the corresponding continuous semantic vector is $\boldsymbol{u}_m = e(\tilde{\boldsymbol{u}}_m) \in \mathbb{S}^{M-1}$ under the embedding map of Definition 3.1.2.1.

**Definition 3.2.1** (Transformer as a Non-Linear Time-Varying VAR Process): Combining the preceding ingredients, the generative trajectory of the Transformer is governed, at each step $i = n + 1, ..., N$, by the autoregressive rule

$$\boldsymbol{u}_i = \boldsymbol{u}_{m_i}, \quad m_i \sim \mathbb{P}_\mathrm{T}(\cdot; \mathcal{P}), \tag{54}$$

on $\{1, ..., M\}$, where $\mathbb{P}_\mathrm{T}(\cdot; \mathcal{P})$ is the Boltzmann distribution of Definition 2.3.1, with score vector

$$\boldsymbol{z}_i = \boldsymbol{U}\Psi_\mathrm{FFN}(\tilde{\boldsymbol{v}}_i) \in \mathbb{R}^M, \quad \boldsymbol{U} = [\boldsymbol{u}_1, ..., \boldsymbol{u}_M]^T \in \mathbb{R}^{M\times M}, \tag{55}$$

so that

$$\mathbb{P}_\mathrm{T}(m; \mathcal{P}) = \exp\left(z_{i,m}/T\right) / \sum_{l=1}^{M} \exp\left(z_{i,l}/T\right) \tag{56}$$

with $T$ the thermodynamic temperature regulating decoding stochasticity. In the low-temperature limit $T \to 0$, sampling collapses to greedy / ML decoding:

$$\boldsymbol{u}_i = \boldsymbol{u}_{m_i^*}, \quad m_i^* = \arg \max_{m\in\{1,...,M\}} \left(\boldsymbol{u}_m^T \Psi_\mathrm{FFN}(\tilde{\boldsymbol{v}}_i)\right). \tag{57}$$

Lifting this one level further, the mechanism above admits the generalized template

$$\boldsymbol{u}_i = \boldsymbol{u}_{m_i}, \quad m_i \sim \mathbb{P}_\mathrm{T}(\cdot; \mathcal{P}), \quad \boldsymbol{z}_i = \boldsymbol{U}\Psi\left(\sum_{j=1}^{i-1} \boldsymbol{W}_{ij}\boldsymbol{u}_j\right), \quad i = n + 1, ..., N, \tag{58}$$

where $\Psi$ is an arbitrary high-dimensional non-linear map, $\boldsymbol{W}_{ij} \in \mathbb{R}^{d\times d}$ is a time-varying transition matrix, and $\mathbb{P}_\mathrm{T}(\cdot; \mathcal{P})$ is the Boltzmann distribution built from the scores $\boldsymbol{z}_i$ (cf. Definition 3.2.1). Stochasticity again enters only through the sampling step, so the deterministic skeleton is a **non-linear time-varying VAR-type recurrence**. The canonical Transformer is the specialization $\boldsymbol{W}_{ij} = \pi_{ij}\boldsymbol{W}_\mathrm{v}$ (a shared linear map $\boldsymbol{W}_\mathrm{v}$ gated by the time-varying Attention scalar $\pi_{ij}$) and $\Psi = \Psi_\mathrm{FFN}$.

Consequently, factorizations or structural simplifications of the transition matrix $\boldsymbol{W}_{ij}$ give rise to new Attention architectures. For example, state-space models such as Mamba / Mamba2 implement a linear-recurrence simplification of the transition dynamics [47], [48]; as discussed in Section 2, however, the linear structure limits their expressiveness on directional, asymmetric semantic dependencies, which has motivated the industry move toward hybrid Attention designs.

The non-linear map $\Psi$ is an equally rich design axis. One line of work replaces the canonical FFN with modern continuous-state Hopfield networks [65]. Retrieval-augmented generation (RAG) [98], backed by large-scale vector databases and knowledge graphs, can be read as replacing $\Psi$ with a query-dependent map $\Psi_i$ that additionally consumes retrieved context at each step, thereby augmenting the parametric memory with up-to-date factual and domain-specific information.

As established in Section 2, the inferential ceiling of LLMs, in the sense of Pearl's hierarchy, is at the level of human Granger-causal inference, instantiated through the autoregressive next-token prediction task. Viewed purely as high-dimensional time-series analysis, Granger-causality detection aims to decipher directional, temporally correlated statistical dependencies between coupled sequences, a methodology that has spread across physics, neuroscience, network science, macroeconomics, and quantitative finance.

Specialising Definition 2.4.1 to LLMs, define the full prefix $\sigma$-algebra $\mathcal{I}_{i-1} = \sigma(\boldsymbol{u}_{1:i-1})$ (containing both the prompt $\boldsymbol{u}_{1:n}$ and the autoregressively generated continuation $\boldsymbol{u}_{n+1:i-1}$) and the prompt-excised sub-$\sigma$-algebra $\mathcal{I}_{i-1}^{-} = \sigma(\boldsymbol{u}_{n+1:i-1})$. Under the informative-prompt assumption of Section 2, there exists a Borel event $\mathcal{A} \in \mathcal{B}(\mathbb{S}^{M-1})$ such that

$$\mathbb{P}(\boldsymbol{u}_i \in \mathcal{A} \mid \mathcal{I}_{i-1}) \neq \mathbb{P}(\boldsymbol{u}_i \in \mathcal{A} \mid \mathcal{I}_{i-1}^{-}) \tag{59}$$

on a set of positive probability. From a temporal-dynamics standpoint, the input/output token trajectories synthesized by LLMs therefore fit the formal Granger-causal paradigm, reaffirming the thesis of Section 2: **in the sense of Pearl's hierarchy, the apparent "reasoning" of LLMs reduces to near-human-level Granger-causal inference, emerging entirely from the minimalist objective of next-token prediction**.

## 3.3. ELBO of the Transformer

Following the notation of Section 2, fix a deterministic input prompt of length $n$, written as the token sequence $S_{1:n}$ with semantic-vector realization $\boldsymbol{s}_{1:n}$ (via the embedding map of Definition 3.1.2.1). For the generation steps $i = n+1, ..., N$, let $U_i$ denote the $i$-th token autoregressively produced by the LLM, with realization $\boldsymbol{u}_i \in \mathbb{S}^{M-1}$; for notational uniformity across the prompt/generation boundary, we set $U_i = S_i$ and $\boldsymbol{u}_i = \boldsymbol{s}_i$ for $i = 1, ..., n$. Let $U_i^{\mathrm{h}}$ denote the ground-truth target token mandated by human preference at step $i$, with corresponding semantic vector $\boldsymbol{u}_i^{\mathrm{h}} \in \mathbb{S}^{M-1}$.

The generative mechanism of the Transformer formalized in Definition 3.2.1 can be re-read through variational inference. Following Kim et al. [99], introduce a discrete latent variable $J$ on the prefix-position set $\{1, ..., i-1\}$, and re-interpret the Attention coefficient $\pi_{ij}$ as the conditional probability

$$\mathbb{Q}(J = j \mid U_{1:i-1}; \mathcal{P}) = \pi_{ij}, \quad j = 1, ..., i-1, \tag{60}$$

where $\mathcal{P}$ is the parameter configuration of Definition 2.3.1. The deterministic Attention aggregate of Definition 3.2.1 then rewrites as a variational expectation,

$$\tilde{\boldsymbol{v}}_i = \sum_{j=1}^{i-1} \pi_{ij} \boldsymbol{W}_{\mathrm{v}} \boldsymbol{u}_j = \mathbb{E}_{\mathbb{Q}(J \mid U_{1:i-1}; \mathcal{P})}[\boldsymbol{W}_{\mathrm{v}} \boldsymbol{u}_J], \tag{61}$$

so the autoregressive rule of Definition 3.2.1 becomes

$$\boldsymbol{u}_i = \boldsymbol{u}_{m_i}, \quad m_i \sim \mathbb{P}_{\mathrm{T}}(\cdot; \mathcal{P}), \quad \boldsymbol{z}_i = \boldsymbol{U} \Psi_{\mathrm{FFN}}\Big(\mathbb{E}_{\mathbb{Q}(J \mid U_{1:i-1}; \mathcal{P})}[\boldsymbol{W}_{\mathrm{v}} \boldsymbol{u}_J]\Big), \tag{62}$$

for $i = n + 1, ..., N$. Applying variational inference [50] to this latent-variable formulation gives the ELBO analysis that follows.

**Theorem 3.3.1** (Evidence Lower Bound (ELBO) of the Transformer): Under the latent-variable variational posterior $\mathbb{Q}\left(J \mid U^{\mathrm{h}}_{1:i-1}; \mathcal{P}\right) = \pi_{ij}$, for each inference step $i = n + 1, ..., N$ the Transformer training objective satisfies

$$\max_{\mathcal{P}} \log \mathbb{P}\left(U^{\mathrm{h}}_i \mid U^{\mathrm{h}}_{1:i-1}; \mathcal{P}\right) \geq \max_{\mathcal{P}} \mathrm{ELBO}\left(\mathbb{Q}\left(J \mid U^{\mathrm{h}}_{1:i-1}; \mathcal{P}\right)\right), \tag{63}$$

with equality iff $\mathbb{Q}\left(J \mid U^{\mathrm{h}}_{1:i-1}; \mathcal{P}\right) = \mathbb{P}\left(J \mid U^{\mathrm{h}}_i, U^{\mathrm{h}}_{1:i-1}; \mathcal{P}\right)$. Maximizing the ELBO therefore maximizes the conditional log-likelihood exactly when the Attention-weight posterior coincides with the true posterior over latent positions.

*Proof*: Pre-training minimizes the cross-entropy between the human reference distribution $\mathbb{P}^{\mathrm{h}}_i = \mathbb{P}\left(U^{\mathrm{h}}_i \mid U^{\mathrm{h}}_{1:i-1}\right)$ and the model distribution $\mathbb{P}_i(\mathcal{P}) = \mathbb{P}\left(U_i \mid U^{\mathrm{h}}_{1:i-1}; \mathcal{P}\right)$:

$$\min_{\mathcal{P}} H\left(\mathbb{P}^{\mathrm{h}}_i \wedge \mathbb{P}_i(\mathcal{P})\right) = \max_{\mathcal{P}} \mathbb{E}_{\mathbb{P}^{\mathrm{h}}_i}[\log \mathbb{P}_i(\mathcal{P})], \quad i = n + 1, ..., N. \tag{64}$$

Pointwise, each training example places a Dirac mass on its ground-truth token, so the per-sample objective reduces to maximum-likelihood estimation of the observed $U^{\mathrm{h}}_i$:

$$\max_{\mathcal{P}} \mathbb{E}_{\mathbb{P}^{\mathrm{h}}_i}[\log \mathbb{P}_i(\mathcal{P})] = \max_{\mathcal{P}} \log \mathbb{P}\left(U^{\mathrm{h}}_i \mid U^{\mathrm{h}}_{1:i-1}; \mathcal{P}\right). \tag{65}$$

Introducing the latent position $J$ (marginalization over the prefix index set) and an arbitrary variational posterior $\mathbb{Q}\left(J \mid U^{\mathrm{h}}_{1:i-1}; \mathcal{P}\right)$ supported on $\{1, ..., i - 1\}$,

$$\begin{aligned} \log \mathbb{P}\left(U^{\mathrm{h}}_i \mid U^{\mathrm{h}}_{1:i-1}; \mathcal{P}\right) &= \log \sum_{j=1}^{i-1} \mathbb{P}\left(U^{\mathrm{h}}_i, J = j \mid U^{\mathrm{h}}_{1:i-1}; \mathcal{P}\right) \\ &= \log \mathbb{E}_{\mathbb{Q}}\left[\frac{\mathbb{P}\left(U^{\mathrm{h}}_i, J \mid U^{\mathrm{h}}_{1:i-1}; \mathcal{P}\right)}{\mathbb{Q}\left(J \mid U^{\mathrm{h}}_{1:i-1}; \mathcal{P}\right)}\right]. \end{aligned} \tag{66}$$

Jensen's inequality (log is concave) yields the ELBO,

$$\begin{aligned} \log \mathbb{P}\left(U^{\mathrm{h}}_i \mid U^{\mathrm{h}}_{1:i-1}; \mathcal{P}\right) &\geq \mathbb{E}_{\mathbb{Q}}\left[\log \frac{\mathbb{P}\left(U^{\mathrm{h}}_i, J \mid U^{\mathrm{h}}_{1:i-1}; \mathcal{P}\right)}{\mathbb{Q}\left(J \mid U^{\mathrm{h}}_{1:i-1}; \mathcal{P}\right)}\right] \\ &= \mathrm{ELBO}\left(\mathbb{Q}\left(J \mid U^{\mathrm{h}}_{1:i-1}; \mathcal{P}\right)\right). \end{aligned} \tag{67}$$

Applying Bayes's chain rule

$$\mathbb{P}\left(U^{\mathrm{h}}_i, J \mid U^{\mathrm{h}}_{1:i-1}; \mathcal{P}\right) = \mathbb{P}\left(J \mid U^{\mathrm{h}}_i, U^{\mathrm{h}}_{1:i-1}; \mathcal{P}\right) \mathbb{P}\left(U^{\mathrm{h}}_i \mid U^{\mathrm{h}}_{1:i-1}; \mathcal{P}\right) \tag{68}$$

inside the expectation and noting that $\mathbb{P}\left(U^{\mathrm{h}}_i \mid U^{\mathrm{h}}_{1:i-1}; \mathcal{P}\right)$ is independent of $J$, the ELBO decomposes as

$$\begin{aligned}&\mathrm{ELBO}\big(\mathbb{Q}\big(J \mid U^{\mathrm{h}}_{1:i-1}; \mathcal{P}\big)\big)\\&= \log \mathbb{P}\big(U^{\mathrm{h}}_i \mid U^{\mathrm{h}}_{1:i-1}; \mathcal{P}\big) - D_{\mathrm{KL}\left(\mathbb{Q}\left(J \mid U^{\mathrm{h}}_{1:i-1}; \mathcal{P}\right) \,\|\, \mathbb{P}\left(J \mid U^{\mathrm{h}}_i, U^{\mathrm{h}}_{1:i-1}; \mathcal{P}\right)\right)}.\end{aligned} \tag{69}$$

Since $D_{\mathrm{KL}} \geq 0$ with equality iff $\mathbb{Q} = \mathbb{P}\big(J \mid U^{\mathrm{h}}_i, U^{\mathrm{h}}_{1:i-1}; \mathcal{P}\big)$, the stated bound and equality condition follow. ■

Theorem 3.3.1 offers a variational-inference reading of Transformer training with four concrete implications. **(i) Unified training-inference objective.** The same $\mathbb{Q}\big(J \mid U^{\mathrm{h}}_{1:i-1}; \mathcal{P}\big) = \pi_{ij}$ governing inference-time Attention is the variational posterior that maximizes the ELBO at training time; the softmax normalization of Attention is therefore not a heuristic but a **provably optimal posterior approximation** under the Jensen bound. **(ii) Equality gap = posterior mismatch.** The bound is tight exactly when $\mathbb{Q}$ coincides with the true latent-position posterior $\mathbb{P}\big(J \mid U^{\mathrm{h}}_i, U^{\mathrm{h}}_{1:i-1}; \mathcal{P}\big)$; the residual $D_{\mathrm{KL}}(\mathbb{Q} \,\|\, \mathbb{P}(J \mid \cdot))$ quantifies how far Attention deviates from the Bayes-optimal prefix assignment, giving a rigorous handle on the representational loss induced by softmax Attention in non-realizable regimes. **(iii) Connection to the EBM formulation.** Because $\log \mathbb{P}\big(U^{\mathrm{h}}_i \mid \cdot; \mathcal{P}\big) = -E_{\mathrm{T}}\Big(\tilde{\boldsymbol{u}}_{U^{\mathrm{h}}_i}; \mathcal{P}\Big)/T - \log Z_{\mathrm{T}}(\mathcal{P})$ under Definition 2.3.1, maximizing the ELBO is equivalent to jointly lowering the energy at the human-aligned token and sharpening the variational posterior toward its Bayesian optimum. This reconciles the statistical-physics view of Section 2 with the signal-processing view of this section. **(iv) Design guidance.** Architectural modifications that enrich $\mathbb{Q}$ (e.g., multi-head Attention, structured or sparse Attention masks) can be evaluated, inside this framework, by the single scalar $D_{\mathrm{KL}}(\mathbb{Q} \,\|\, \mathbb{P}(J \mid \cdot))$ — a principled criterion for comparing Attention variants beyond empirical benchmarks.

## 3.4. Alternative Architectures

To reduce the computational cost of the canonical Transformer in both pre-training and autoregressive inference, the AI community has proposed a range of alternative LLM architectures. We analyze the leading candidates within the generalized non-linear time-varying VAR-type framework of Equation 58, evaluating each by the specific choice of transition matrix $\boldsymbol{W}_{ij}$ and non-linear map $\Psi$ it induces.

### 3.4.1. Mamba Series

To avoid the softmax computation in Attention, Mamba / Mamba2 [47], [48] replace self-Attention with a discrete **state-space model (SSM)**. Writing $\boldsymbol{h}_i$ for the hidden state (to avoid collision with the generated-token notation $\boldsymbol{u}_i$ of Definition 3.2.1) and $\boldsymbol{x}_i$ for the input embedding at step $i$, the SSM recursion is

$$\begin{cases}\boldsymbol{h}_i = \boldsymbol{A}_i \boldsymbol{h}_{i-1} + \boldsymbol{B}_i \boldsymbol{x}_i\\ \boldsymbol{y}_i = \boldsymbol{C}_i \boldsymbol{h}_i\end{cases} \tag{70}$$

which unrolls to a **linear time-varying VAR** recurrence [46] :

$$\boldsymbol{h}_i = \sum_{j \leq i} \left( \prod_{k=j+1}^{i} \boldsymbol{A}_k \right) \boldsymbol{B}_j \boldsymbol{x}_j. \tag{71}$$

This corresponds to the specialization of Equation 58 in which the transition matrix absorbs the product $\prod \boldsymbol{A}_k \cdot \boldsymbol{B}_j$ (still time-varying but **linear in the past**, with no bilinear-form gate) and the non-linear map $\Psi$ collapses to the linear readout $\boldsymbol{C}_i$. Linear time-varying VAR processes are mature objects in econometrics and finance [100], [101], and their estimation toolkit is directly transferable to Mamba-family training.

As discussed in Section 2, the absence of a bilinear semantic-relevance gate makes pure Mamba/Mamba2 theoretically less expressive than Transformers on directional dependencies. Still, the architecture motivates the search for models that match Transformer expressiveness at markedly lower compute. Building on the gated-Mamba2 of [102], Qwen3-Next is a production-scale **hybrid-Attention** LLM that interleaves Mamba-style SSM blocks with standard Attention blocks.[6] The residual non-linearity of Transformer — the softmax of the bilinear form $\boldsymbol{u}_{i-1}^T \boldsymbol{J} \boldsymbol{u}_j$ — is exactly what SSMs drop, and is what a hybrid design restores selectively.

Mamba3 [103] pushes this line further through three technical changes: (i) a trapezoidal discretization scheme that removes the need for explicit causal convolutions, (ii) a complex-valued state space enabling improved discrete state-tracking, and (iii) a multi-input multi-output (MIMO) formulation that raises hardware arithmetic intensity without sacrificing decoding latency. Empirically, at the 1.5B-parameter scale, Mamba3 outperforms the closest sub-quadratic competitor by roughly 1.8 points across downstream benchmarks, matches Mamba2's generation quality with half the state dimension, and performs well on synthetic algorithmic reasoning. Its limitations are equally concrete: retrieval quality degrades on very long, unstructured context — an unavoidable consequence of fixed-state compression — the MIMO design inflates pre-training cost, and hybridizing with classical self-Attention introduces nontrivial design trade-offs.

### 3.4.2. LLaDA

LLaDA [49] is a **discrete diffusion** LLM that departs from the canonical autoregressive template. At training time, a mask ratio $\tau \in (0, 1]$ is drawn and a $\tau$-fraction of the tokens $U_{1:n}^0$ in an utterance is replaced by a special mask symbol $[\mathsf{MASK}]$, yielding the corrupted sequence $U_{1:n}^\tau$ (here $U_{1:n}^0$ denotes the uncorrupted ground-truth sequence). The network then predicts, from the unmasked context, the original identity of each masked position. Concretely, LLaDA minimizes the masked-position cross-entropy

$$\mathcal{L}(\mathcal{P}) = -\mathbb{E}_{\tau, U_{1:n}^\tau, U_{1:n}^0} \left[ \frac{1}{\tau} \sum_{i=1}^{n} \mathbf{1}(U_i^\tau = [\mathsf{MASK}]) \log \mathbb{P}(U_i^0 \mid U_{1:n}^\tau; \mathcal{P}) \right], \tag{72}$$

where the indicator $\mathbf{1}(U_i^\tau = [\mathsf{MASK}])$ restricts the sum to masked coordinates, and the $1/\tau$ weighting compensates the expected number $n\tau$ of masked positions and makes the objective an unbiased estimator of the discrete-diffusion ELBO. A non-causal (bidirectional) Transformer is used as the backbone. Despite the discrete-diffusion framing, LLaDA fits the generalized VAR-type recurrence of Equation 58 in its non-causal variant: the per-step aggregate $\sum_{j \neq i} \boldsymbol{W}_{ij} \boldsymbol{u}_j$ replaces the causal $\sum_{j<i}$ of Definition 3.2.1, while the $(\boldsymbol{W}_{ij}, \Psi)$ parameterization is retained.

[6] https://qwen.ai/blog?from=research.latest-advancements-list&id=4074cca80393150c248e508aa62983f9cb7d27cd&

## 3.5. Information Theory and Signal Processing

The mapping between the discrete semantic space and its continuous semantic-vector counterpart mirrors the classical dichotomy between information theory and signal processing. Grounded in probability theory, classical information theory sets out the fundamental limits of information compression, transmission, and storage while setting aside implementation-complexity constraints [15]. Signal processing, by contrast, realizes these abstract quantities in continuous geometric spaces — canonically $\mathbb{R}^n$ or $\mathbb{C}^n$ — and thereby unlocks the continuous-analysis toolkit needed to design tractable algorithms for perception, communication, and memory [104].

Recasting the opening thesis of this section in these terms, the use of vectorized token sequences in LLMs is not a mere engineering heuristic but a deliberate **conversion of probability-centric NLP methodology into high-dimensional signal processing**. Semantic vectorization is the precise meeting point of the two disciplines: continuous geometric optimization is used to operationalize the abstract limits set by information theory.

At a higher level, it is useful to contrast information theory with Turing's theory of computation. Turing's framework is concerned with the discrete, deterministic operational steps required to execute a given algorithm, so its central metric is computational latency, typically measured by time complexity. Shannon's framework, by contrast, is concerned with achievability and converse limits of communication rates and data compression, in a regime where the coding block length is taken to infinity and latency is, by construction, out of scope.

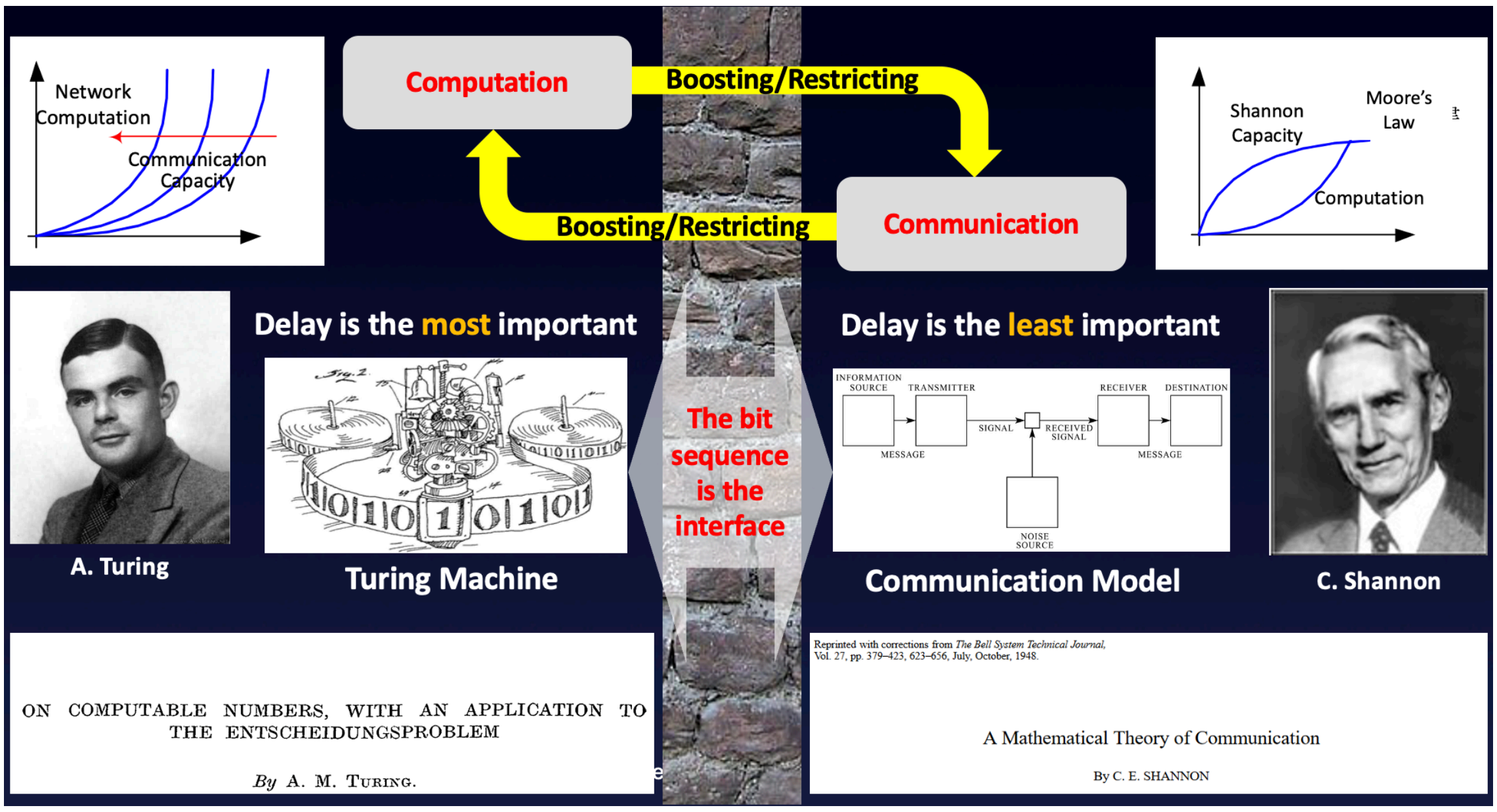

Figure 2: The **BIT** as the shared atomic unit bridging Turing's theory of computation and Shannon's theory of communication — the defining conceptual breakthrough of the Information Epoch.

This theoretical divergence produces a mirror-image duality in engineering practice. Modern communication systems are bounded by available compute and decoding-time budgets; modern computational systems, in turn, are bounded by communication-bandwidth limits across the memory hierarchy, inter-GPU links, and inter-node interconnects. Yet, despite their different roots and objectives, both disciplines rest on the same atomic unit: the **BIT**. We may therefore state it directly: **the BIT is the common atomic substrate bridging the historically distinct worlds of computation and communication.**

This point was made forcefully by Prof. Erdal Arikan during our panel colloquium (§1): **the BIT is the defining conceptual breakthrough of the Information Epoch.** Shannon and Weaver [2] confined classical information theory to the reliable transmission of symbols — a purely syntactic and physical engineering problem — and deliberately abstracted away semantic meaning and pragmatic effectiveness. As the global paradigm shifts toward the AI Epoch, the question becomes whether classical information theory can retain that foundational role. As shown in Section 4, lifting the atomic unit from the Information-Epoch **BIT** to the AI-Epoch **TOKEN** generalizes Shannon's framework into a **Semantic Information Theory** adequate for the generative behavior of modern LLMs.

# 4. Information Theory Perspective

## 4.1. Information Theory and AI

The intersection of information theory and AI is not a modern development; it goes back to AI's origin. Claude Shannon, the founder of information theory, was a principal organizer of the 1956 Dartmouth Summer Research Project on AI [105] — the workshop now recognized as AI's disciplinary birthplace. Shannon's formulation of entropy and statistical uncertainty shaped early AI concretely: it provided quantitative bounds on heuristic search spaces and guided sequential decision-making, as in his 1950 paper on computer chess [106].

In the subsequent decades, information-theoretic methods reshaped natural-language processing. During the 1970s–80s, Frederick Jelinek and his IBM group fused statistical and information-theoretic ideas into the core of speech recognition and NLP [107]. By treating language generation as a stochastic communication process governed by hidden Markov models (HMMs) and $n$-gram distributions, Jelinek led the shift from rule-based expert systems to data-driven statistical NLP, laying the groundwork for modern statistical language modeling.

The cross-pollination was sealed by a mathematical unification of error-correcting coding and probabilistic inference. The **belief propagation (BP)** algorithm introduced by Judea Pearl for Bayesian inference on graphical models [108] was later shown to be the same algorithm as the iterative message-passing decoder for LDPC codes [109]. This established graphical models and message passing as a common algorithmic substrate between information theory and machine learning, and subsequent work has continued in the same vein: **approximate message passing (AMP)** for compressive sensing [110], and the use of **Rényi information dimension** as a fundamental limit in compressed sensing [111].

In the deep-neural-network era, this relationship has deepened further. Information-theoretic principles continue to illuminate the mathematical mechanics and generalization properties of modern machine-learning models [112]; the **information bottleneck** principle, in particular, has become a standard lens on learning dynamics, phase transitions, and representational trade-offs in deep networks [113]. The same machinery is now being brought to bear on the LLM "black box": classical rate-distortion and information-bottleneck formulations have been used to analyze semantic vectorization in large models [114]; autoregressive language models have been repurposed as textual transform coders to improve multimedia compression [115]; rate-distortion theory has been extended with perceptual-quality constraints into the **rate-distortion-perception** function for evaluating generative fidelity [116]; and Transformer Attention has been modeled as an interacting-particle system to characterize its asymptotic clustering behavior [117]. The centrality of massive datasets in LLM pre-training only reinforces information theory's role as a foundational lens for data science [118].

Despite this resurgence, a concrete gap remains: the autoregressive generative mechanics of LLMs — and of the Transformer in particular — have not yet been axiomatized from a rigorous information-theoretic standpoint.

## 4.2. Classical Shannon Information Theory

This section reviews the main theoretical limits and methodological insights of Shannon's classical information theory. The schematic in Figure 3 is Shannon's 1948 depiction of a general communication system [1].

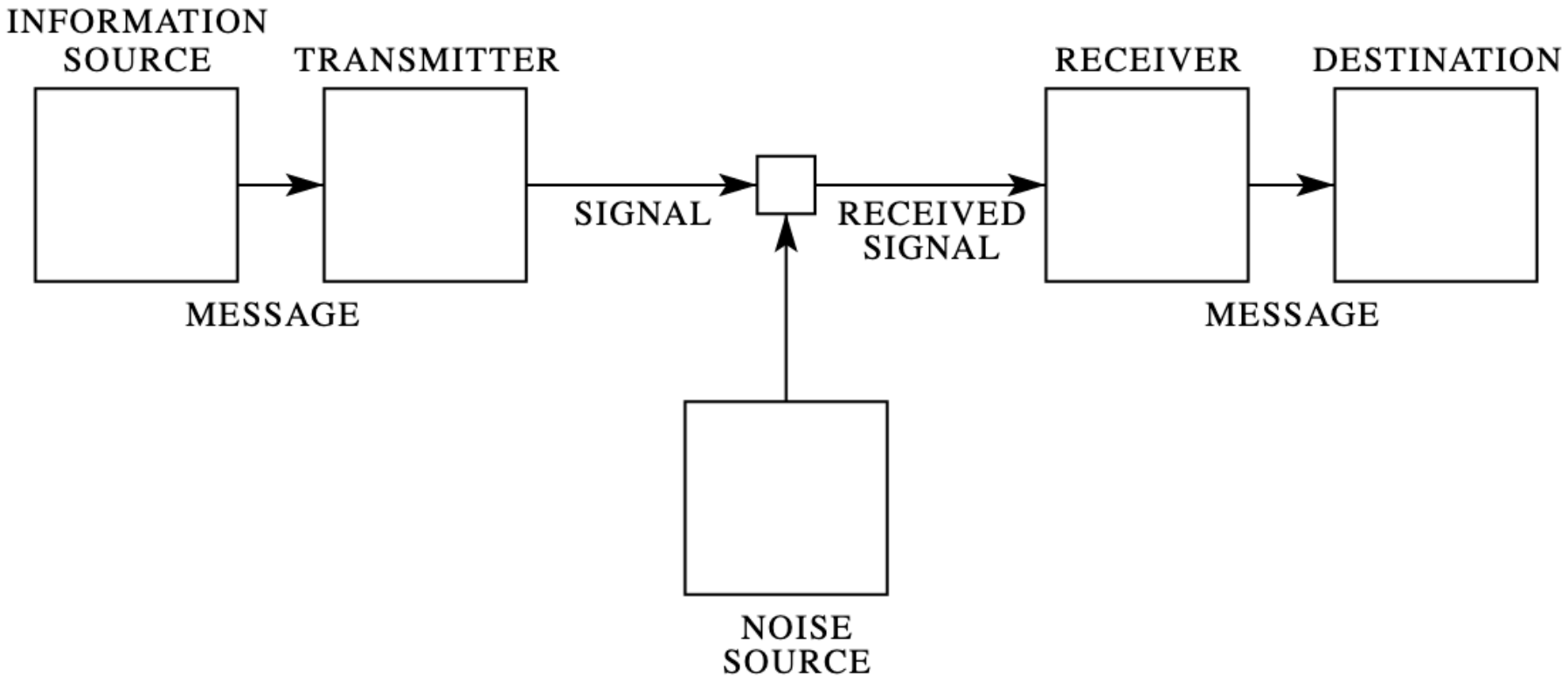

Figure 3: Shannon's canonical schematic of a general communication system.

### 4.2.1. Three Fundamental Theorems of Information Theory

In a classical communication system, the information source generates the data to be transmitted. The role of the source encoder is to map each source symbol to a binary codeword as short as possible, compressing the payload and thereby conserving channel bandwidth and improving transmission efficiency. If the source emits i.i.d. samples of a discrete random variable $S$ with probability mass $\mathbb{P}$ on a finite alphabet $\Omega$ (not to be confused with the LLM vocabulary $\mathcal{U}$ of Definition 3.1.1.1), Shannon showed that the amount of information produced per symbol is quantified by the **entropy**

$$H(S) = -\sum_{s \in \Omega} \mathbb{P}(s) \log \mathbb{P}(s). \tag{73}$$

Shannon's **Source Coding Theorem** then states that $H(S)$ is the asymptotic infimum of the expected codeword length per source symbol among all lossless block codes: no uniquely-decodable code can do strictly better than $H(S)$ on average, and this limit is achievable in the block-length-to-infinity regime.

Physical communication channels are corrupted by stochastic noise, so transmitting raw source symbols leads to errors at the receiver. Reliable digital communication over noisy media had long been an open problem that defied engineering heuristics; Shannon established both its feasibility and an optimal path to it. He modeled the **communication channel** by the conditional transition law

$$\mathbb{P}(Y_{1:n} \mid X_{1:n}) \tag{74}$$

of a discrete memoryless channel (DMC), where $X_{1:n}$ is the channel input sequence and $Y_{1:n}$ is the output sequence of the same length observed at the receiver. To combat noise, Shannon introduced **channel coding**; at the receiver, **maximum-likelihood (ML) decoding**

$$\hat{m} = \arg\max_{m} \mathbb{P}(Y_{1:n} \mid X_{1:n}(m)) \tag{75}$$

minimizes the block error probability $P_e$ over uniform codebooks. If the channel code carries a message set of cardinality $M_c$ (distinct from the LLM vocabulary size $M$ of Definition 3.1.1.1), the **communication rate** is defined as

$$R = \frac{1}{n} \log M_c \quad \text{bits/channel use,} \tag{76}$$

assuming messages are uniformly distributed.

For a DMC, the $n$-th extension of the channel factorizes across time [119]:

$$\mathbb{P}(Y_{1:n} \mid X_{1:n}) = \prod_{k=1}^{n} \mathbb{P}(Y_k \mid X_k). \tag{77}$$

Shannon showed that, for any DMC, the supremum of reliably achievable communication rates $R$ equals the **channel capacity**

$$C = \max_{\mathbb{P}(X)} I(X;Y), \tag{78}$$

where the maximum is taken over all input distributions $\mathbb{P}(X)$ and $I(X;Y)$ is the **mutual information** between channel input $X$ and output $Y$. For $X \in \mathcal{X}$ and $Y \in \mathcal{Y}$ with finite alphabets,

$$I(X;Y) = \sum_{x \in \mathcal{X}} \sum_{y \in \mathcal{Y}} \mathbb{P}(x,y) \log \frac{\mathbb{P}(x,y)}{\mathbb{P}(x)\mathbb{P}(y)}. \tag{79}$$

Shannon proved that, whenever the communication rate $R$ is strictly below $C$, there exist sequences of channel codes whose block error probability satisfies $P_e \to 0$ as the block length $n \to \infty$; conversely, whenever $R > C$, no code sequence can achieve $P_e \to 0$. This result is the **Noisy-Channel Coding Theorem**, and it identifies $C$ as the exact critical rate: an achievable rate below $C$ and an unbreakable barrier above it. Crucially, $C$ is defined as the supremum of $I(X;Y)$ over all input distributions $\mathbb{P}(X)$, and the capacity-achieving prior $\mathbb{P}^*(X)$ specifies the statistical structure any capacity-approaching code must asymptotically emulate. As developed below, lifting this classical DMC formulation to an autoregressive setting yields the central information-theoretic concept we need for LLMs: **directed information**.

Shannon's third contribution is the **Source-Channel Separation Theorem**: provided $H(S) < C$, a communication system that separately performs source coding (down to rate $H(S)$) and channel coding (up to rate $C$) is asymptotically optimal, incurring no loss relative to any joint design in the block-length-to-infinity limit; conversely, $H(S) > C$ implies $P_e$ is bounded away from zero. This decoupling turned the discipline into two largely independent specialties — source compression and channel transmission — each of which can be advanced without detailed knowledge of the other.

#### 4.2.2. Methodological Implications

Shannon's 1948 treatise is a landmark for using rigorous mathematics to resolve an engineering problem that had resisted empirical attack (**transcending physical empiricism through mathematical axiomatization**). Its deepest methodological move was to refuse engagement with any particular hardware or transient technology. Instead, Shannon asked a counterfactual question: if a perfectly reliable communication system existed, what functions would it have to perform and what mathematical properties would it have to satisfy? This is a textbook **top-down** deduction — the necessary conditions for reliable information transmission are derived at the level of runtime behavior, yielding a framework that governs the architecture of every concrete communication system built afterwards.

In the specific setting of channel coding and decoding, Shannon's treatise answers three questions:

1. **What is reliability in digital communication, mathematically?**
   - Shannon defined reliability as asymptotically error-free transmission, and by bringing the axioms of probability and mathematical statistics into telecommunications he introduced the analytical toolkit — bounded error probabilities, error exponents, ML decoding, and joint typicality decoding — now standard in the field.
2. **What is the universal mathematical model of a communication channel?**
   - Shannon represented the physical channel by a conditional transition law. The abstraction is crucial: regardless of whether the hardware exists today or will be invented later, its stochastic uncertainty can be captured by transition probabilities. The model is **implementation-agnostic**, and hence universal.
   - From a mathematical standpoint this is an instance of the **probabilistic method** [120].
3. **What are the performance metrics for a communication system?**
   - Shannon showed that the reliable communication rate satisfies $R \leq C = \max_{\mathbb{P}(X)} I(X;Y)$: the channel capacity is an upper bound on achievable rates, and it is attained by the optimal input distribution $\mathbb{P}^*(X)$. Mutual information, in turn, is itself a special case of Kullback-Leibler (KL) divergence, which measures the statistical distance between two distributions: for discrete $\mathbb{P}, \mathbb{Q}$ on $\mathcal{X}$ with $\mathbb{P} \ll \mathbb{Q}$,

$$D_{\mathrm{KL}}(\mathbb{P} \parallel \mathbb{Q}) = \sum_{x \in \mathcal{X}} \mathbb{P}(x) \log \frac{\mathbb{P}(x)}{\mathbb{Q}(x)}, \tag{80}$$

so that

$$I(X;Y) = D_{\mathrm{KL}}(\mathbb{P}(X,Y) \parallel \mathbb{P}(X)\mathbb{P}(Y)). \tag{81}$$

The **cross-entropy** loss ubiquitous in modern machine learning admits a parallel decomposition in terms of entropy and KL divergence: using $H(\mathbb{P} \wedge \mathbb{Q}) := -\mathbb{E}_{\mathbb{P}}[\log \mathbb{Q}]$,

$$H(\mathbb{P} \wedge \mathbb{Q}) = H(\mathbb{P}) + D_{\mathrm{KL}}(\mathbb{P} \parallel \mathbb{Q}). \tag{82}$$

Since the entropy $H(\mathbb{P})$ of the target distribution is constant in the model parameters, minimizing empirical cross-entropy is equivalent to minimizing $D_{\mathrm{KL}}(\mathbb{P} \parallel \mathbb{Q})$.

## 4.3. Information-Theoretic Measures for LLMs

This section formalizes the information-theoretic measures we will need for the autoregressive dynamics of LLMs. It combines a classical tool — the **rate-distortion function** — with two measures tailored to temporal and asymmetric dependence: **directed information** and **directed information density**.

#### 4.3.1. Rate-Distortion Function

Lossy compression — reconstructing source symbols approximately rather than exactly — underlies most large-scale data pipelines for image, video, and audio. Shannon introduced the **rate-distortion function** to characterize its ultimate limits [1]. Formally, one seeks the infimum rate $R$ (in bits/symbol) at which the decompressed sequence can be reconstructed with expected distortion at most a prescribed tolerance $D$. Let $d : \mathcal{S} \times \mathcal{S} \to \mathbb{R}_+ \cup \{0\}$ be a per-letter distortion measure, and extend it to sequences by the block average

$$d\left(S_{1:n}, \hat{S}_{1:n}\right) = \frac{1}{n} \sum_{k=1}^{n} d\left(S_k, \hat{S}_k\right), \tag{83}$$

where $S_{1:n}$ is the original source sequence and $\hat{S}_{1:n}$ is its lossy reconstruction.

**Definition 4.3.1.1** (Rate-Distortion Function): For a stationary ergodic source, the **rate-distortion function** $R(D)$ is the infimum of achievable compression rates at distortion level at most $D$:

$$R(D) = \lim_{n\to\infty} \frac{1}{n} \inf_{\mathbb{P}\left(\hat{S}_{1:n} \mid S_{1:n}\right) : \mathbb{E}\left[d\left(S_{1:n}, \hat{S}_{1:n}\right)\right] \leq D} I\left(S_{1:n}; \hat{S}_{1:n}\right), \tag{84}$$

where the infimum is over all conditional distributions $\mathbb{P}\left(\hat{S}_{1:n} \mid S_{1:n}\right)$ that — together with the source law $\mathbb{P}(S_{1:n})$ — yield a joint distribution satisfying the expected-distortion constraint $\mathbb{E}\left[d\left(S_{1:n}, \hat{S}_{1:n}\right)\right] \leq D$.

Following Shannon's methodology, this definition abstracts the entire encoder/decoder pipeline into a single conditional law $\mathbb{P}\left(\hat{S}_{1:n} \mid S_{1:n}\right)$. The variational problem is to find the conditional law that minimizes the mutual information between source and reconstruction subject to the expected-distortion constraint; here the mutual information serves as the rate required to describe $\hat{S}_{1:n}$ within distortion $D$, not as a correlation penalty.

It is worth emphasizing that the objective is $I\left(S_{1:n}; \hat{S}_{1:n}\right)$ rather than the marginal entropy $H\left(\hat{S}_{1:n}\right)$. Using the identity

$$I\left(S_{1:n}; \hat{S}_{1:n}\right) = H(S_{1:n}) - H\left(S_{1:n} \mid \hat{S}_{1:n}\right), \tag{85}$$

the conditional entropy $H\left(S_{1:n} \mid \hat{S}_{1:n}\right)$ is exactly the residual uncertainty in $S_{1:n}$ after observing the reconstruction, i.e., the information deliberately discarded to meet the distortion tolerance $D$. An optimal lossy scheme therefore preserves only the component of $S_{1:n}$ correlated with $\hat{S}_{1:n}$, which is why $I\left(S_{1:n}; \hat{S}_{1:n}\right)$ is the natural objective. The alternative $H\left(\hat{S}_{1:n}\right)$ is the rate required to describe $\hat{S}_{1:n}$ losslessly — a different problem — and one has the elementary inequality

$$I\left(S_{1:n}; \hat{S}_{1:n}\right) \leq H\left(\hat{S}_{1:n}\right). \tag{86}$$

Because controlled information loss in lossy compression is dual to channel-induced noise, rate-distortion theory is structurally dual to channel coding; for a comprehensive treatment we refer to Berger's monograph [52].

Closed-form evaluation of $R(D)$ is rarely available, and its numerical computation is expensive because the $n$-letter variational problem has $|\hat{\mathcal{S}}|^n$ parameters. The **Blahut-Arimoto (BA) algorithm** [121], [122] addresses this by alternating minimization in the single-letter (discrete memoryless source) setting. Our group has recently proposed an optimal-transport-based algorithm that accelerates this numerical evaluation [123].

### 4.3.2. Directed Information

The concept of **directed information** was introduced by James Massey, the 1988 IEEE Claude Shannon Award laureate [51]. In his 1990 paper, Massey argued that the canonical DMC factorization in Ash's classical textbook [119] cannot capture the causal dynamics of **feedback**. Massey noted, as a historical remark, that Shannon devoted his first Shannon Lecture at the 1973 IEEE International Symposium on Information Theory to feedback channels — a choice that reflected how little the classical theory had to say about closed-loop communication at the time.

Massey's first observation was that a genuinely memoryless channel (which we call the **Massey-DMC**) must satisfy the Markov property

$$\mathbb{P}(Y_n \mid X_{1:n}, Y_{1:n-1}) = \mathbb{P}(Y_n \mid X_n), \tag{87}$$

i.e., the output $Y_n$ depends only on the current input $X_n$, independent of the rest of the input history and all past outputs. Separately, a discrete channel used **without feedback** (the **Massey-DNFC**) is one in which the input law satisfies

$$\mathbb{P}(X_n \mid X_{1:n-1}, Y_{1:n-1}) = \mathbb{P}(X_n \mid X_{1:n-1}), \tag{88}$$

so that $X_n$ is shaped by its own history $X_{1:n-1}$ but carries no influence from past outputs $Y_{1:n-1}$; equivalently, the input generator runs in open loop. Massey then showed that the two properties are jointly necessary and sufficient for the Ash-style factorization: the channel is simultaneously a Massey-DMC and a Massey-DNFC if and only if its $n$-th extension satisfies

$$\mathbb{P}(Y_{1:n} \mid X_{1:n}) = \prod_{k=1}^{n} \mathbb{P}(Y_k \mid X_k) \quad \text{(Ash-DMC)}. \tag{89}$$

As used in the classical channel-capacity theorem, the mutual information $I(X;Y)$ measures the statistical dependence between the input and output of an Ash-DMC, but it is symmetric in its arguments and therefore cannot capture the directional (causal) dependence present in a Massey-DMC with feedback. Building on Marko's earlier work on bidirectional communication [124], Massey introduced **directed information** to quantify precisely this causal, asymmetric dependence between input and output sequences.

**Definition 4.3.2.1** (Directed Information): The **directed information** from an input sequence $X_{1:n}$ to an output sequence $Y_{1:n}$ is

$$I(X_{1:n} \to Y_{1:n}) = \sum_{k=1}^{n} I(X_{1:k}; Y_k \mid Y_{1:k-1}). \tag{90}$$

Massey established the following basic properties of this quantity:

**Theorem 4.3.2.1** (Properties of Directed Information):
1. For any discrete channel with input $X_{1:n}$ and output $Y_{1:n}$,

$$I(X_{1:n} \to Y_{1:n}) \leq I(X_{1:n}; Y_{1:n}), \tag{91}$$

with equality if and only if the channel is used without feedback (i.e., it is a Massey-DNFC).

2. For a Massey-DMC,

$$I(X_{1:n} \to Y_{1:n}) \leq \sum_{k=1}^{n} I(X_k; Y_k), \tag{92}$$

with equality if and only if the input sequence $X_{1:n}$ is i.i.d. (equivalently, since the channel is memoryless, the outputs $Y_{1:n}$ are i.i.d.).

Massey also gave a precise definition of causality in communication systems. He required that, for a causal channel, the output obey the conditional independence

$$\mathbb{P}(Y_n \mid Y_{1:n-1}, X_{1:n}, S_{1:m}) = \mathbb{P}(Y_n \mid Y_{1:n-1}, X_{1:n}), \tag{93}$$

where $S_{1:m}$ is the full source output (with $m$ potentially exceeding $n$, i.e., including source symbols that have not yet been transmitted). The condition says that **the source acts on the channel output $\boldsymbol{Y_n}$ only through the current input $\boldsymbol{X_{1:n}}$ and the past outputs $\boldsymbol{Y_{1:n-1}}$**: $Y_n$ cannot look into the source directly, in particular not into its future. We refer to this property as **Massey causality**. Under this axiom, Massey established a third property of directed information:

**Theorem 4.3.2.2** (Directed Information and Massey Causality): Let $X_{1:n}$ and $Y_{1:n}$ be the input and output sequences of a Massey-causal discrete channel, and let $S_{1:m}$ denote the source output. Then

$$I(S_{1:m}; Y_{1:n}) \leq I(X_{1:n} \to Y_{1:n}). \tag{94}$$

Chaining Theorem 4.3.2.2 with Property 2 of Theorem 4.3.2.1 (which applies whenever the channel is also a Massey-DMC) gives

$$I(S_{1:m}; Y_{1:n}) \le I(X_{1:n} \to Y_{1:n}) \le \sum_{k=1}^{n} I(X_k; Y_k), \tag{95}$$

from which one recovers the classical (and initially counter-intuitive) fact that **noiseless feedback does not increase the capacity of a DMC**.

The results above show that directed information extends the reach of (symmetric) mutual information to a broad class of channels with memory and feedback. For decades, however, mainstream information-theoretic practice stayed within the classical Ash-DMC framework — which has no place for causal feedback — and directed information received little attention outside specialist work. For a full treatment and modern applications, we refer to Kramer's PhD thesis [125] and Tsur's survey [126].

By reversing the time ordering in Definition 4.3.2.1, we introduce the **backward directed information**.[7]

**Definition 4.3.2.2** (Backward Directed Information): The **backward directed information** from the reverse-indexed input sequence $X_{n:1}$ to the forward-indexed output sequence $Y_{1:n}$ is the sum of step-wise conditional mutual informations between the **future** inputs $X_{k+1:n}$ and the current output $Y_k$, conditioned on the past outputs $Y_{1:k-1}$:

$$I(X_{n:1} \to Y_{1:n}) = \sum_{k=1}^{n} I(X_{k+1:n}; Y_k \mid Y_{1:k-1}). \tag{96}$$

This is exactly the objective maximized by the information-theoretically optimal semantic encoder of Definition 3.1.4.1: the family $\{f_k\}$ producing contextual embeddings $\boldsymbol{U}_k = f_k(X_{1:k})$ that satisfies

$$I^*(X_{n:1} \to \boldsymbol{U}_{1:n}) = \sum_{k=1}^{n} \max_{f_k} I(X_{k+1:n}; \boldsymbol{U}_k \mid \boldsymbol{U}_{1:k-1}) \tag{97}$$

maximizes, at each step $k$, the conditional mutual information between the future $X_{k+1:n}$ and the embedding $\boldsymbol{U}_k$ — i.e., the backward-directed-information contribution of step $k$.

### 4.3.3. Directed Information Density

**Information density** was introduced by Dobrushin in 1959 [128] (some historical accounts attribute an earlier idea to Pinsker). Strassen gave the first rigorous asymptotic analysis in his 1962 thesis [129].

[7] The name "backward" is inspired by backward stochastic differential equations, developed by Shige Peng [127] as the foundation for his nonlinear expectation theory, a framework parallel to Kolmogorov's axiomatic probability.

**Definition 4.3.3.1** (Information Density): Let $X$ and $Y$ be discrete random variables on alphabets $\mathcal{X}$ and $\mathcal{Y}$ with joint mass function $\mathbb{P}(x, y)$ and marginals $\mathbb{P}(x)$, $\mathbb{P}(y)$. The **information density** of a realization pair $(x, y)$ is the log-likelihood ratio

$$\imath(x; y) = \log \frac{\mathbb{P}(x, y)}{\mathbb{P}(x)\mathbb{P}(y)} = \log \frac{\mathbb{P}(y \mid x)}{\mathbb{P}(y)}. \tag{98}$$

Information density is a scalar random variable whose law is induced by the joint distribution $\mathbb{P}(X, Y)$; averaging it recovers the Shannon mutual information,

$$I(X; Y) = \mathbb{E}_{\mathbb{P}(X,Y)}[\imath(X; Y)] = \sum_{x \in \mathcal{X}} \sum_{y \in \mathcal{Y}} \mathbb{P}(x, y) \imath(x; y). \tag{99}$$

It has become a standard tool for analyzing communication limits in the non-asymptotic, finite-blocklength regime [112].

Extending information density to the causally ordered temporal setting gives the **directed information density**.

**Definition 4.3.3.2** (Directed Information Density): For realizations $x_{1:n}$ and $y_{1:n}$ of stochastic input and output sequences, the **directed information density** from $x_{1:n}$ to $y_{1:n}$ is the time-indexed sum

$$\imath(x_{1:n} \to y_{1:n}) = \sum_{k=1}^{n} \imath(x_{1:k}; y_k \mid y_{1:k-1}), \tag{100}$$

where the causally conditioned log-likelihood ratio at step $k$ is

$$\imath(x_{1:k}; y_k \mid y_{1:k-1}) = \log \frac{\mathbb{P}(y_k \mid y_{1:k-1}, x_{1:k})}{\mathbb{P}(y_k \mid y_{1:k-1})}. \tag{101}$$

Like its non-causal counterpart, directed information density is a random variable along the sequence, and its expectation under the joint distribution recovers the directed information,

$$\begin{aligned} I(X_{1:n} \to Y_{1:n}) &= \mathbb{E}_{\mathbb{P}(X_{1:n}, Y_{1:n})}[\imath(X_{1:n} \to Y_{1:n})] \\ &= \sum_{k=1}^{n} \mathbb{E}_{\mathbb{P}(X_{1:k}, Y_{1:k})}[\imath(X_{1:k}; Y_k \mid Y_{1:k-1})] = \sum_{k=1}^{n} I(X_{1:k}; Y_k \mid Y_{1:k-1}). \end{aligned} \tag{102}$$

## 4.4. Information-Theoretic Abstraction of LLMs

Classical information theory analyzes communication systems from a **runtime** standpoint, an angle that turns out to be well suited to the mathematical first principles of LLMs. Our goal is a theoretical framework for LLMs that is decoupled from any particular architectural instance: even when the field moves beyond today's Transformer, such a framework should remain applicable. This methodological stance is shared by 2024 Turing laureate Richard Sutton, who in proposing the Oak architecture argued that progress toward artificial general intelligence (AGI) requires a clean separation between design-time algorithms and runtime dynamics [130].

Mirroring Shannon's three questions for reliable communication (Section 4), we pose three analogous questions for LLMs:

1. **What are "semantics" for LLMs, mathematically?**
2. **What is a universal information-theoretic model of LLMs, independent of specific architectures?**
3. **What are the right performance metrics for the inferential capabilities of LLMs?**

The first is answered by the signal-processing framework of Section 3, so the remainder of this section focuses on the second and third.

#### 4.4.1. LLM as a Channel with Feedback and State

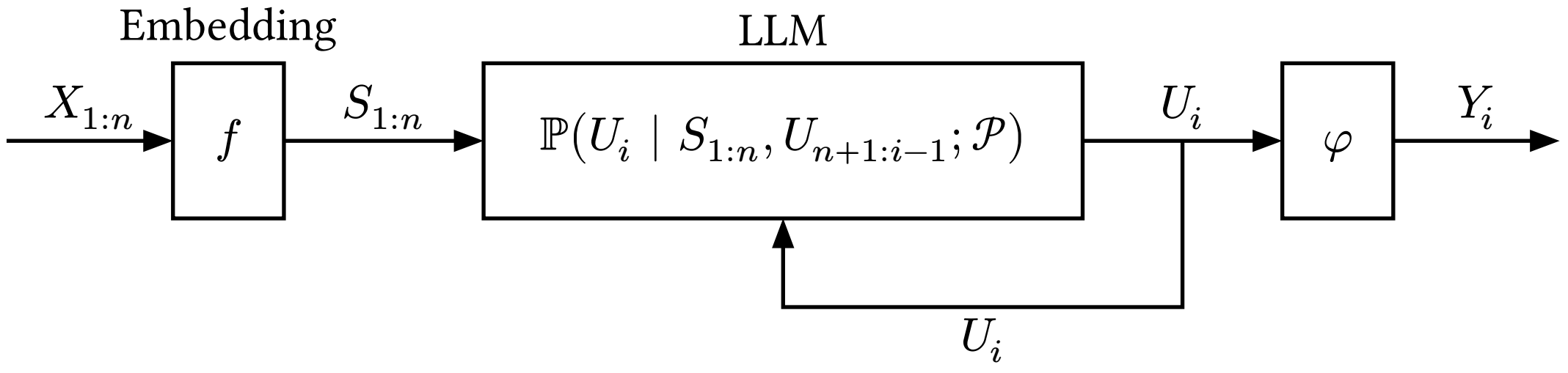

Figure 4: Information-theoretic abstraction of an LLM as a causal communication channel with state and an autoregressive feedback loop.

As illustrated in Figure 4, let $X_{1:n}$ be the discrete prompt sequence, with indices $1 \leq n < i \leq N$, $n \in \mathbb{N}$. The semantic encoder $f$ maps this sequence to its continuous semantic-vector representation $\boldsymbol{S}_{1:n}$. Conditioned on $\boldsymbol{S}_{1:n}$ and on the previously generated internal states $\boldsymbol{U}_{n+1:i-1}$, the autoregressive engine produces the next internal state $\boldsymbol{U}_i$. The generative dynamics of the LLM are thus captured by the conditional law

$$\mathbb{P}(\boldsymbol{U}_i \mid \boldsymbol{S}_{1:n}, \boldsymbol{U}_{n+1:i-1}; \mathcal{P}), \tag{103}$$

where $\mathcal{P} = \{\boldsymbol{W}_{\mathrm{q}}, \boldsymbol{W}_{\mathrm{k}}, \boldsymbol{W}_{\mathrm{v}}, \mathcal{W}, \mathcal{B}\}$ collects all trainable parameters. A decoder $\varphi$ (not a functional inverse of $f$, but a sampling / arg-max map, cf. Definition 3.2.1) turns $\boldsymbol{U}_i$ into the emitted discrete token $Y_i$. Viewed through an information-theoretic lens, the LLM is exactly a **causal communication channel with internal memory state and autoregressive feedback**. The abstraction is architecture-agnostic (Transformers, state-space models, and so on), so studying it yields design guidance for next-generation generative architectures. The abstraction also highlights a pair of structural contrasts between classical Shannon communication and LLM generation:

- **Classical communication.** The goal is asymptotically error-free reconstruction at the receiver. Since $X$ and $Y$ live at different ends of the channel and have different alphabets, the natural object is not some distance between their marginals but their statistical dependence. Shannon's **mutual information** $I(X;Y) = D_{\mathrm{KL}}(\mathbb{P}(X,Y) \parallel \mathbb{P}(X)\mathbb{P}(Y))$ measures how far the joint distribution deviates from independence; the larger it is, the more reliably $X$ can be decoded from $Y$. Maximizing $I(X;Y)$ over admissible input distributions $\mathbb{P}(X)$ gives the channel capacity. **In machine-learning language, the loss functional driving the classical communication problem is mutual information, a form of KL divergence.**

- **LLM generation.** The goal is to approximate the conditional law of the human continuation $Y_{n+1:N}$ given a prompt $X_{1:n}$. Since human-annotated corpora provide direct access to samples from this ground-truth distribution, training can use KL divergence directly to fit the model distribution to the empirical target. Because the entropy of the target is constant in the model parameters (cf. Section 4.2.2), **the variational optimization of LLMs reduces to minimizing cross-entropy, an equivalent surrogate for KL divergence**.

This contrast supports the central thesis of Section 1: once the atomic unit of classical information theory is lifted from the **BIT** to the **TOKEN**, the generative mechanics of LLMs admit a clean, unified information-theoretic formalization.

### 4.4.2. The Theoretical Principles of LLM Training via Semantic Information Theory

Since the LLM is a stateful communication channel with internal memory and autoregressive feedback, the directed information from the vectorized prompt $\boldsymbol{S}_{1:n}$ to the generated sequence $\boldsymbol{U}_{n+1:N}$ quantifies the causal transmission of semantic information through the model. Let $\boldsymbol{U}^{\mathrm{h}}_{n+1:N}$ denote the human ground-truth continuation in response to $\boldsymbol{S}_{1:n}$. For notational uniformity across the prompt/generation boundary, set $\boldsymbol{U}_i = \boldsymbol{S}_i$ for $i = 1, ..., n$. Write

$$\mathbb{P}^{\mathrm{h}}_i = \mathbb{P}\big(\boldsymbol{U}^{\mathrm{h}}_i \mid \boldsymbol{U}^{\mathrm{h}}_{1:i-1}\big), \quad \mathbb{P}_i(\mathcal{P}) = \mathbb{P}(\boldsymbol{U}_i \mid \boldsymbol{U}_{1:i-1}; \mathcal{P}), \quad i = n+1, ..., N, \tag{104}$$

for the step-wise conditional distributions of the human reference and the parameterized model.

**Definition 4.4.2.1** (Directed Rate-Distortion Function for Pre-training): The **directed rate-distortion function** for LLM pre-training is the constrained variational problem

$$R_{\mathrm{pre}}(D) = \frac{1}{N} \min_{\mathcal{P}: \frac{1}{N}\sum_{i=n+1}^{N} D_{\mathrm{KL}}(\mathbb{P}^{\mathrm{h}}_i \,\|\, \mathbb{P}_i(\mathcal{P})) \le D} I(\boldsymbol{S}_{1:n} \to \boldsymbol{U}_{n+1:N}; \mathcal{P}), \tag{105}$$

with both normalizations taken per sequence position. The directed information admits the chain-rule decomposition

$$I(\boldsymbol{S}_{1:n} \to \boldsymbol{U}_{n+1:N}; \mathcal{P}) = \sum_{i=n+1}^{N} I(\boldsymbol{S}_{1:n}; \boldsymbol{U}_i \mid \boldsymbol{U}_{n+1:i-1}; \mathcal{P}), \tag{106}$$

a prompt-to-generation extension of Massey's directed information (Definition 4.3.2.1) in which the input block $\boldsymbol{S}_{1:n}$ is fixed and the output grows autoregressively.

The formulation above uses KL divergence as the distortion metric for consistency with the earlier definitions. The function $R_{\mathrm{pre}}(D)$ is the theoretical infimum for end-to-end pre-training performance, and admits two useful readings.

- **Minimum causal rate.** $R_{\mathrm{pre}}(D)$ is the minimum rate of causal information that must flow from the prompt $\boldsymbol{S}_{1:n}$ to the generation $\boldsymbol{U}_{n+1:N}$ for the step-wise model distribution to stay within KL distance $D$ of the human reference. Tracking $R_{\mathrm{pre}}(D)$ during training can, in principle, reveal phase transitions and emergent capabilities of LLMs.

- **Hallucination mitigation via directed information bottleneck.** Minimizing the directed-information term inside $R_{\mathrm{pre}}(D)$ acts as a regularizer in the spirit of the **information bottleneck** [131]: it compresses the representation $\boldsymbol{U}_i$ by suppressing prompt-to-representation correlations that are not needed to match the human reference, a principled route to suppressing hallucinations. This motivates the Lagrangian pre-training loss

$$\mathcal{L}(\mathcal{P}) = I(\boldsymbol{S}_{1:n}; \boldsymbol{U}_i \mid \boldsymbol{U}_{n+1:i-1}; \mathcal{P}) + \lambda H\left(\mathbb{P}_i^{\mathrm{h}} \wedge \mathbb{P}_i(\mathcal{P})\right), \quad \forall i = n+1, ..., N, \tag{107}$$

where $\lambda > 0$ is a Lagrange multiplier trading rate against distortion.

A direct calculation yields the following characterization of the pre-training optimum.

**Theorem 4.4.2.1** (Asymptotic Limit of Pre-training): Assuming global convergence under the cross-entropy objective, the zero-distortion pre-training rate equals the per-position directed information of the human reference:

$$R_{\mathrm{pre}}(0) = \frac{1}{N} I\left(\boldsymbol{S}_{1:n} \to \boldsymbol{U}_{n+1:N}^{\mathrm{h}}\right). \tag{108}$$

*Proof*: The standard identity

$$H\left(\mathbb{P}_i^{\mathrm{h}} \wedge \mathbb{P}_i(\mathcal{P})\right) = H\left(\mathbb{P}_i^{\mathrm{h}}\right) + D_{\mathrm{KL}}\left(\mathbb{P}_i^{\mathrm{h}} \parallel \mathbb{P}_i(\mathcal{P})\right), \quad i = n+1, ..., N, \tag{109}$$

combined with the fact that $H\left(\mathbb{P}_i^{\mathrm{h}}\right)$ does not depend on $\mathcal{P}$, shows that cross-entropy minimization is equivalent to KL minimization:

$$\min_{\mathcal{P}} H\left(\mathbb{P}_i^{\mathrm{h}} \wedge \mathbb{P}_i(\mathcal{P})\right) \Leftrightarrow \min_{\mathcal{P}} D_{\mathrm{KL}}\left(\mathbb{P}_i^{\mathrm{h}} \parallel \mathbb{P}_i(\mathcal{P})\right), \quad i = n+1, ..., N. \tag{110}$$

At a global minimizer $\mathcal{P}^{\dagger}$ the model distribution matches the human reference,

$$\mathbb{P}_i\left(\mathcal{P}^{\dagger}\right) = \mathbb{P}\left(\boldsymbol{U}_i \mid \boldsymbol{U}_{1:i-1}; \mathcal{P}^{\dagger}\right) = \mathbb{P}\left(\boldsymbol{U}_i^{\mathrm{h}} \mid \boldsymbol{U}_{1:i-1}^{\mathrm{h}}\right) = \mathbb{P}_i^{\mathrm{h}}, \quad i = n+1, ..., N, \tag{111}$$

so $D_{\mathrm{KL}}\left(\mathbb{P}_i^{\mathrm{h}} \parallel \mathbb{P}_i\left(\mathcal{P}^{\dagger}\right)\right) = 0$ and $D = 0$. Because the joint laws of $\left(\boldsymbol{S}_{1:n}, \boldsymbol{U}_{n+1:N}\right)$ under $\mathcal{P}^{\dagger}$ and of $\left(\boldsymbol{S}_{1:n}, \boldsymbol{U}_{n+1:N}^{\mathrm{h}}\right)$ under the human reference coincide, each step-wise conditional mutual information matches the corresponding human-reference quantity, and expanding the directed information via Definition 4.3.2.1 yields

$$\begin{aligned} I\left(\boldsymbol{S}_{1:n} \to \boldsymbol{U}_{n+1:N}; \mathcal{P}^{\dagger}\right) &= \sum_{i=n+1}^{N} I\left(\boldsymbol{S}_{1:n}; \boldsymbol{U}_i \mid \boldsymbol{U}_{n+1:i-1}; \mathcal{P}^{\dagger}\right) \\ &= \sum_{i=n+1}^{N} I\left(\boldsymbol{S}_{1:n}; \boldsymbol{U}_i^{\mathrm{h}} \mid \boldsymbol{U}_{n+1:i-1}^{\mathrm{h}}\right) = I\left(\boldsymbol{S}_{1:n} \to \boldsymbol{U}_{n+1:N}^{\mathrm{h}}\right). \end{aligned} \tag{112}$$

Dividing by $N$ completes the proof. ■

**Remark 4.4.2.1** (Information Geometry and Pre-training): Consider two parameter points $\mathcal{P}$ and $\mathcal{P}'$ in the parameter manifold, and the corresponding model distributions $\mathbb{P}_i(\mathcal{P})$ and $\mathbb{P}_i(\mathcal{P}')$ of step $i$. Following Amari's **information geometry** [132], the $(k, l)$ -entry of the **Fisher information matrix (FIM)** is the Hessian of the KL divergence,

$$[\mathcal{I}(\mathcal{P})]_{kl} = \frac{\partial^2}{\partial[\mathcal{P}']_k \partial[\mathcal{P}']_l} D_{\mathrm{KL}}(\mathbb{P}_i(\mathcal{P}) \parallel \mathbb{P}_i(\mathcal{P}'))\Bigg|_{\mathcal{P}'=\mathcal{P}}, \tag{113}$$

equivalent here to the Hessian of the cross-entropy $H(\mathbb{P}_i(\mathcal{P}) \wedge \mathbb{P}_i(\mathcal{P}'))$ in $\mathcal{P}'$ (the two differ only by the constant $H(\mathbb{P}_i(\mathcal{P}))$). Thus the FIM captures the Riemannian curvature of the loss landscape in $\mathcal{P}$. Preconditioning the Euclidean gradient by $\mathcal{I}(\mathcal{P})^{-1}$ yields **natural gradient descent**, which takes steps in this information-geometric metric rather than the Euclidean one. Because exact FIM inversion is infeasible at the scale of modern networks, the **Kronecker-Factored Approximate Curvature (K-FAC)** algorithm [133] is the standard practical approximation.

Thus, at ideal convergence the pre-training parameters settle at $\mathcal{P}^\dagger$, and the directed information flow of the pre-trained model approaches the upper bound $I\big(\boldsymbol{S}_{1:n} \to \boldsymbol{U}^{\mathrm{h}}_{n+1:N}\big)$ set by the human reference.

While pre-training optimizes next-token prediction, its resulting long-horizon generations may drift from human values or preferences. Post-training alignment addresses this by evaluating the full generated sequence against human preferences, typically via **Reinforcement Learning from Human Feedback (RLHF)** [32].

RLHF introduces a learned scalar **reward model** $w\big(\boldsymbol{S}_{1:n}, \boldsymbol{U}_{n+1:N}\big) \in \mathbb{R}$ that scores the generated sequence $\boldsymbol{U}_{n+1:N}$ given the prompt $\boldsymbol{S}_{1:n}$. This leads to the following definition.

**Definition 4.4.2.2** (Directed Rate-Reward Function for Post-training): The **directed rate-reward function** for LLM post-training is the constrained variational problem

$$R_{\mathrm{post}}(W) = \frac{1}{N} \inf_{\mathcal{P}^\dagger : \mathbb{E}[w(\boldsymbol{S}_{1:n}, \boldsymbol{U}_{n+1:N})] \geq W} I\big(\boldsymbol{S}_{1:n} \to \boldsymbol{U}_{n+1:N}; \mathcal{P}^\dagger\big), \tag{114}$$

where the infimum is over the post-pre-training parameter configurations $\mathcal{P}^\dagger$ (the class initialized from the pre-training result and then refined by alignment), and the expectation is taken jointly over prompts $\boldsymbol{S}_{1:n}$ drawn from the training distribution and generations $\boldsymbol{U}_{n+1:N} \sim \mathbb{P}\big(\cdot \mid \boldsymbol{S}_{1:n}; \mathcal{P}^\dagger\big)$. The minimizer is $\mathcal{P}^\ddagger$, the post-training-aligned parameter configuration of Section 2.

Written as a Lagrangian in $\mathcal{P}^\dagger$, the post-training objective becomes

$$\mathcal{L}\big(\mathcal{P}^\dagger\big) = I\big(\boldsymbol{S}_{1:n} \to \boldsymbol{U}_{n+1:N}; \mathcal{P}^\dagger\big) - \lambda \mathbb{E}\big[w\big(\boldsymbol{S}_{1:n}, \boldsymbol{U}_{n+1:N}\big)\big], \tag{115}$$

with $\lambda > 0$ trading directed-information rate against expected reward; the minimizer is denoted $\mathcal{P}^{\ddagger}$. Structurally, this is in the same spirit as the reward-plus-regularization objective used in **Direct Preference Optimization (DPO)** [134].

By Theorem 4.4.2.1, pre-training drives the per-position directed information of the model toward $\frac{1}{N} I(\boldsymbol{S}_{1:n} \to \boldsymbol{U}^{\mathrm{h}}_{n+1:N})$ — the causal semantic information that a human generator transfers from prompt $\boldsymbol{S}_{1:n}$ to response $\boldsymbol{U}^{\mathrm{h}}_{n+1:N}$ (per position). Post-training then moves the parameters from $\mathcal{P}^{\dagger}$ to $\mathcal{P}^{\ddagger}$ so that the generated sequence $\boldsymbol{U}_{n+1:N}$ also respects human preferences encoded by $w$. Combining this picture with the directed-information–Granger-causality correspondence of Section 2.4 yields the following corollary.

**Corollary 4.4.2.1** (Granger-Causal Equivalence of Aligned LLMs): After full pre-training and post-training alignment, the autoregressive next-token predictions of the LLM approximate, in the sense of Pearl's hierarchy, the Granger-causal structure of human generation subject to the preference constraints encoded by the reward model $w$.

### 4.4.3. Semantic Information Theory Principles for Inference

At inference time, the aligned LLM — with parameters fixed at the post-training optimum $\mathcal{P}^{\ddagger}$ — generates a specific output sequence $\boldsymbol{U}_{n+1:N}$ in response to a given prompt $\boldsymbol{S}_{1:n}$. Unlike training, which optimizes an expectation over an ensemble of sequences, inference operates on a single sample path. We therefore track the (realization-level) directed information density of Definition 4.3.3.2 rather than its expectation.

**Definition 4.4.3.1** (Semantic Information Flow for Inference): The **semantic information flow** during autoregressive inference is the cumulative directed information density from the prompt $\boldsymbol{S}_{1:n}$ to the generated prefix $\boldsymbol{U}_{n+1:i}$:

$$\imath(\boldsymbol{S}_{1:n} \to \boldsymbol{U}_{n+1:i}; \mathcal{P}^{\ddagger}) = \sum_{k=n+1}^{i} \imath(\boldsymbol{S}_{1:n}; \boldsymbol{U}_k \mid \boldsymbol{U}_{n+1:k-1}; \mathcal{P}^{\ddagger}) \tag{116}$$

for $i = n + 1, ..., N$.

Autoregressive inference terminates at a dedicated end-of-sequence token $\lhd$, so the length $N$ of the generated sequence is the stopping time $N = \inf\{i \geq n + 1 : \boldsymbol{U}_i = e(\lhd)\}$ with respect to the generation filtration $\sigma(\boldsymbol{U}_{n+1}, \boldsymbol{U}_{n+2}, ...)$, where $e(\lhd) \in \mathbb{S}^{M-1}$ is the semantic-vector embedding of $\lhd$ in the sense of Definition 3.1.2.1. The following theorem characterizes the stochastic dynamics of semantic information flow under this stopping rule.

**Theorem 4.4.3.1** (Sub-Martingale Property of Semantic Information Flow): Let $\mathcal{G}_{i-1} = \sigma(\imath(\boldsymbol{S}_{1:n} \to \boldsymbol{U}_{n+1:k}; \mathcal{P}^\ddagger) : n+1 \le k \le i-1)$ be the natural filtration generated by the semantic information flow up to step $i-1$. The cumulative semantic information flow is a discrete-time sub-martingale with respect to $\mathcal{G}_{i-1}$ for $i = n+1, ..., N$.

*Proof*: By Definition 4.3.3.2, the flow admits the one-step additive decomposition

$$\begin{aligned}&\imath(\boldsymbol{S}_{1:n} \to \boldsymbol{U}_{n+1:i}; \mathcal{P}^\ddagger)\\ &= \imath(\boldsymbol{S}_{1:n} \to \boldsymbol{U}_{n+1:i-1}; \mathcal{P}^\ddagger) + \imath(\boldsymbol{S}_{1:n}; \boldsymbol{U}_i \mid \boldsymbol{U}_{n+1:i-1}; \mathcal{P}^\ddagger),\end{aligned} \tag{117}$$

with the step-wise conditional information density

$$\imath(\boldsymbol{S}_{1:n}; \boldsymbol{U}_i \mid \boldsymbol{U}_{n+1:i-1}; \mathcal{P}^\ddagger) = \log \frac{\mathbb{P}(\boldsymbol{U}_i \mid \boldsymbol{U}_{n+1:i-1}, \boldsymbol{S}_{1:n}; \mathcal{P}^\ddagger)}{\mathbb{P}(\boldsymbol{U}_i \mid \boldsymbol{U}_{n+1:i-1}; \mathcal{P}^\ddagger)}. \tag{118}$$

The first term on the right is $\mathcal{G}_{i-1}$-measurable (it is one of the generators of $\mathcal{G}_{i-1}$) and so passes through conditional expectation. For the second term, write $\mathcal{F}_{i-1} = \sigma(\boldsymbol{U}_{n+1:i-1})$ for the token-level filtration; since $\mathcal{G}_{i-1} \subseteq \mathcal{F}_{i-1}$, the tower property gives

$$\begin{aligned}&\mathbb{E}[\imath(\boldsymbol{S}_{1:n}; \boldsymbol{U}_i \mid \boldsymbol{U}_{n+1:i-1}; \mathcal{P}^\ddagger) \mid \mathcal{G}_{i-1}]\\ &= \mathbb{E}[D_{\mathrm{KL}}(\mathbb{P}(\boldsymbol{U}_i \mid \boldsymbol{U}_{n+1:i-1}, \boldsymbol{S}_{1:n}; \mathcal{P}^\ddagger) \parallel \mathbb{P}(\boldsymbol{U}_i \mid \boldsymbol{U}_{n+1:i-1}; \mathcal{P}^\ddagger)) \mid \mathcal{G}_{i-1}],\end{aligned} \tag{119}$$

where we used the standard identity that the $\mathcal{F}_{i-1}$-conditional expectation of the log-likelihood ratio in $\boldsymbol{U}_i$ equals the KL divergence between the two step-wise conditional laws. By Gibbs's inequality $D_{\mathrm{KL}} \ge 0$ pointwise, the right-hand side is non-negative almost surely, and therefore

$$\mathbb{E}[\imath(\boldsymbol{S}_{1:n} \to \boldsymbol{U}_{n+1:i}; \mathcal{P}^\ddagger) \mid \mathcal{G}_{i-1}] \ge \imath(\boldsymbol{S}_{1:n} \to \boldsymbol{U}_{n+1:i-1}; \mathcal{P}^\ddagger), \tag{120}$$

which is the sub-martingale inequality. ■

Given this sub-martingale property, we apply the **Doob decomposition**: with respect to the filtration $\mathcal{G}_{i-1}$ of Theorem 4.4.3.1, the information flow admits a unique additive factorization

$$\imath(\boldsymbol{S}_{1:n} \to \boldsymbol{U}_{n+1:i}; \mathcal{P}^\ddagger) = M_i + A_i, \tag{121}$$

where $A_i$ is $\mathcal{G}$-predictable and non-decreasing, and $M_i$ is a $\mathcal{G}$-martingale. The compensator is given by the cumulative one-step conditional expectations,

$$A_i = \sum_{j=n+1}^{i} \mathbb{E}[\imath(\boldsymbol{S}_{1:n} \to \boldsymbol{U}_{n+1:j}; \mathcal{P}^\ddagger) - \imath(\boldsymbol{S}_{1:n} \to \boldsymbol{U}_{n+1:j-1}; \mathcal{P}^\ddagger) \mid \mathcal{G}_{j-1}], \tag{122}$$

with $A_n = 0$; the martingale part is then $M_i = \imath(\boldsymbol{S}_{1:n} \to \boldsymbol{U}_{n+1:i}; \mathcal{P}^\ddagger) - A_i$. To quantify the stochastic deviation in $M_i$, introduce its predictable quadratic variation

$$V_i = \sum_{j=n+1}^{i} \mathbb{E}\Big[\big(M_j - M_{j-1}\big)^2 \mid \mathcal{G}_{j-1}\Big]. \tag{123}$$

The next corollary is then an immediate consequence of Freedman's martingale tail inequality [135].

**Corollary 4.4.3.1** (Concentration Bound for Semantic Information Flow): Assume the martingale differences are almost surely bounded, $|M_j - M_{j-1}| \le c$ a.s. for some $c > 0$. Then for any $\alpha, \beta > 0$,

$$\mathbb{P}(M_i \ge \alpha, V_i \le \beta) \le \exp\left(-\frac{\alpha^2}{2(\beta + c\alpha)}\right). \tag{124}$$

The two assumptions in Corollary 4.4.3.1 play different roles and are both necessary. $V_i$ controls the **average** conditional variance and gives the Gaussian-like regime of the bound for small $\alpha$; the uniform increment bound $|M_j - M_{j-1}| \le c$ controls the **worst-case** jump size and gives the Poisson-type correction $c\alpha$ in the denominator, preventing heavy-tailed spikes from escaping the concentration envelope. In our LLM setting, the martingale increment is the centred log-likelihood ratio $\imath\big(\boldsymbol{S}_{1:n}; \boldsymbol{U}_j \mid \boldsymbol{U}_{n+1:j-1}; \mathcal{P}^{\ddagger}\big) - \mathbb{E}\big[\cdot \mid \mathcal{G}_{j-1}\big]$; since the Boltzmann distribution of Definition 2.3.1 at any finite temperature $T > 0$ assigns every vocabulary token a probability bounded away from 0 and 1, the log-likelihood ratio is itself bounded, and $|M_j - M_{j-1}| \le c$ holds with some $c = c(M, B, T)$ depending only on the vocabulary size $M$, the energy bound $B$ of Theorem 2.3.1.1, and the temperature $T$. The assumption in Corollary 4.4.3.1 is therefore automatic under the setting of this paper.

Combining the sub-martingale property with the **Optional Stopping Theorem** ([136]) at the bounded stopping time $N$ gives a chain-rule-style inequality for the expected information flow.

**Corollary 4.4.3.2** (Optional Stopping of Semantic Information Flow): Provided $N$ is almost-surely bounded,

$$I\big(\boldsymbol{S}_{1:n} \to \boldsymbol{U}_{n+1:N}; \mathcal{P}^{\ddagger}\big) \ge I\big(\boldsymbol{S}_{1:n}; \boldsymbol{U}_{n+1}; \mathcal{P}^{\ddagger}\big). \tag{125}$$

Equivalently, the cumulative directed information up to the halting time lower-bounds the single-step mutual information between the prompt and the first generated token.

Despite the combinatorial complexity of discrete token-sequence distributions and the difficulty of modelling long-range dependencies, **directed information density enjoys clean mathematical properties**: the autoregressive dynamics of LLMs is therefore amenable to the standard toolkit of martingale theory. In the spirit of Shannon capacity (the supremum of mutual information over input distributions), these corollaries motivate an analogous upper bound for inference, presented next.

**Definition 4.4.3.2** (Semantic Information Capacity for Inference): In the spirit of Shannon capacity, and in contrast to the deterministic-prompt setting used elsewhere in this paper, we model the prompt as a random variable with distribution $\mathbb{P}(\boldsymbol{S}_{1:n})$ and define the **semantic information capacity**

$$C_{\text{sem}}(W) = \sup_{\mathbb{P}(\boldsymbol{S}_{1:n}) : \mathbb{E}[w(\boldsymbol{S}_{1:n}, \boldsymbol{U}_{n+1:N})] \geq W} I\big(\boldsymbol{S}_{1:n} \to \boldsymbol{U}_{n+1:N}; \mathcal{P}^{\ddagger}\big), \tag{126}$$

where the supremum is over prompt distributions satisfying the expected reward constraint $\mathbb{E}[w] \geq W$, and the expectation is taken jointly over $\boldsymbol{S}_{1:n} \sim \mathbb{P}(\boldsymbol{S}_{1:n})$ and $\boldsymbol{U}_{n+1:N} \sim \mathbb{P}\big(\cdot \mid \boldsymbol{S}_{1:n}; \mathcal{P}^{\ddagger}\big)$.

Definition 4.4.3.2 says that, at inference time under alignment constraints, the directed semantic information can be further enlarged by optimizing the prompt distribution. This gives an information-theoretic justification for **harness engineering** (also known as **prompt engineering** or **in-context engineering**): better prompts increase the directed semantic information transferred to the model, and thereby the quality of its responses.

### 4.4.4. Estimation of Directed Information

Despite its elegance, directed information is computationally expensive to estimate in high-dimensional state spaces. When the joint distribution is known in closed form, Naiss and Permuter extended the Blahut-Arimoto algorithm to compute directed-information capacity [137], exploiting its concavity in the causal input distribution together with a dynamic-programming recursion. In practice the distribution is almost never known, and estimation must proceed from samples; the dominant sample-based surrogate is **transfer entropy** [138], introduced by the statistical-physics community to quantify Granger-style causal influence between coupled stochastic time series.

**Definition 4.4.4.1** (Transfer Entropy): Let $X_{1:n}$ and $Y_{1:n}$ be two interacting discrete-time stochastic processes. The transfer entropy from the source $X$ to the destination $Y$ at an arbitrary temporal index $i$ is the conditional mutual information

$$T_i(X \to Y) = I(X_{i-L:i-1}; Y_i \mid Y_{i-L:i-1}), \tag{127}$$

where $L \in \mathbb{N}^{+}$ is the finite temporal horizon of Markovian influence.

Transfer entropy is the finite-lag, strictly-past counterpart of each summand in Definition 4.3.2.1: the latter conditions on the full prefix $X_{1:k}$ — including the current source symbol $X_k$ — whereas the former uses a length-$L$ sliding window $X_{i-L:i-1}$ that strictly precedes the destination token $Y_i$. In the parametric case of Gaussian VAR processes, transfer entropy is moreover mathematically isomorphic to Granger causality [139]; for a comprehensive account of the isomorphisms connecting Granger causality and directed information, we refer the reader to Amblard and Michel's survey [140].

The finite-lag truncation is not a mere technical convenience: it matches the operating regime of modern LLMs. A deployed Transformer always conditions on a bounded context window of size $W_{\text{ctx}}$ (e.g., $W_{\text{ctx}} \in \{2^{13}, \ldots, 2^{20}\}$ tokens), so each autoregressive step uses at most $W_{\text{ctx}}$ past tokens. Setting $L \leq W_{\text{ctx}}$ in Definition 4.4.4.1 aligns the two regimes exactly: the Transformer's causal attention computes, token by token, a truncated $W_{\text{ctx}}$-lag instance of the transfer-entropy summand, so directed-information estimation reduces to a data task the Transformer is already solving during next-token training. Conversely, every increase of $W_{\text{ctx}}$ (long-context Transformers) shifts the estimate closer to the full-prefix directed information, providing an information-theoretic reading of why long-context LLMs yield sharper causal estimates.

This alignment is precisely what the neural-estimation literature has progressively uncovered. Belghazi et al. first introduced the **Mutual Information Neural Estimator (MINE)** based on the Donsker-Varadhan variational form of KL divergence [141], [142]; Permuter's group extended it with recurrence to obtain the **Directed Information Neural Estimator (DINE)** for temporally dependent processes [143]; most recently, the same group's **TREET** (TRansfer Entropy Estimation via Transformers) [144] recasts transfer-entropy estimation as an autoregressive prediction problem, making explicit that **the generative dynamics of Transformers and the machinery of directed information share the same underlying structure**.

This two-layer structure is Shannon's own pattern. In classical communication theory, **channel capacity** is defined at the $n \to \infty$ limit of finite-blocklength achievable rates, while practical codes (Polar, LDPC, Turbo, etc.) operate at finite $n$ and only approach capacity as $n$ grows. Exactly the same relation holds here: **full-prefix directed information** is defined at the $W_{\text{ctx}} \to \infty$ limit, while $W_{\text{ctx}}$-truncated transfer entropy — and hence TREET — is its realizable counterpart at a bounded context window. This is why we retain directed information as the primary theoretical object throughout this paper: its asymptotic full-history definition plays the same conceptual role for LLM generation that Shannon capacity plays for communication, and long-context scaling of Transformers can be read as closing the gap between the two.

## 4.5. The Causal Ceiling of LLMs: Granger Prediction without Pearl Intervention

Throughout Section 2 and Section 3 we have built toward a single thesis: the cognitive substance of LLM reasoning is, at base, the computational realization of near-human-level Granger causal inference — an emergent capability produced by an architecturally minimal objective, autoregressive next-token prediction. This identification immediately invites a philosophical rejoinder: does Granger-style predictive dependence — operationalized here through directed information and its transfer-entropy realization (Definition 4.4.4.1) — constitute **genuine** causation, or merely temporally ordered correlation?

Judea Pearl, pioneer of Bayesian networks and 2011 Turing Laureate, has leveled a sustained critique against Granger causality. In his view, Granger's formulation conflates temporal precedence with causal structure and has thereby misled a generation of empirical work: from the structural perspective, Granger causality is not a causal mechanism but **statistical correlation endowed with temporal ordering**. Pearl further insists on an epistemological axiom — observational data, without explicit structural-model assumptions, cannot yield genuine causal conclusions. The resulting dichotomy can be summarized as:

- **Granger causality:** rooted in **empiricism**, driven entirely by data via autoregressive prediction and historical temporal correlation.
- **Pearl causality:** rooted in **causal structuralism**, organized around **interventions** and **counterfactual** reasoning under explicit causal-model assumptions.

More precisely, Pearl organizes causal epistemology into a three-tier "Ladder of Causation":

- **Level 1: Association (Seeing).** Passive observation of statistical correlations in empirical data, formalized by the ordinary conditional probability
$$\mathbb{P}(Y \mid X). \tag{128}$$
- **Level 2: Intervention (Doing).** Active perturbation of the system via the $\mathrm{do}(\cdot)$ operator, which evaluates the distribution of $Y$ after a structural intervention fixes $X$ to a chosen value:
$$\mathbb{P}(Y \mid \mathrm{do}(X)). \tag{129}$$
- **Level 3: Counterfactuals (Imagining).** Having observed the factual pair $\{X = x, Y = y\}$, one asks what $Y$ **would have been** under an alternative antecedent $X = x'$:
$$\mathbb{P}(Y_{x'} \mid X = x, Y = y), \tag{130}$$
where $Y_{x'}$ denotes the potential outcome of $Y$ under the hypothetical assignment $X = x'$.

Pearl's central theorem within this framework states that **a lower rung cannot answer questions at a higher rung without exogenous causal assumptions that are themselves unidentifiable from observational data alone** [41], [145]. Viewed through this prism, Granger causality — despite its temporal autoregressive structure — is confined to Level 1: it is the ceiling of what purely data-driven prediction can attain. By the same theorem, an LLM trained exclusively on passive Level-1 corpora cannot, by its structural mechanics alone, execute genuine Level-2 interventions or Level-3 counterfactual reasoning.

Recent integrations of deliberative algorithms — reinforcement learning combined with Monte Carlo Tree Search (MCTS), for example — into the LLM inference pipeline [146], [147] have markedly improved apparent reasoning ability. But under the structural constraint of a fixed autoregressive generative model, these search procedures can at most asymptotically imitate the **syntactic patterns** of intervention and counterfactual reasoning that are already latent in human corpora. An LLM can produce token sequences that read as genuine interventions or counterfactuals precisely because it has internalized the linguistic distributions of human causal narratives; yet at the epistemological level this remains data-driven predictive mimicry rather than structural causal inference [148]. Seen in this light, the value of human-AI collaboration lies in the human's capacity to inject exogenous causal assumptions — assumptions unidentifiable from the raw textual manifold — thereby providing the structural scaffolding that harnesses the LLM's predictive bandwidth as an engine for human intellectual productivity.

# 5. Epilogue and Conclusions

At the outset, a clarification is in order: the provocative title of this treatise — *Forget BIT, It is All about TOKEN* — is not meant to deprecate the BIT. On the contrary, this work proceeds from the conviction that the **BIT** remains the paramount intellectual achievement of the Information Epoch. This sentiment echoes a remark made by Prof. Erdal Arikan — inventor of the 5G Polar code and recipient of the 2019 Shannon Award — during a roundtable in early 2023, a remark that catalyzed the present search for an atomic substrate of the AI Epoch as fundamental as the BIT once was. **The BIT bridged computation and communication** — two disciplines with distinct theoretical foundations, yet inextricably bound in mutual constraint and facilitation. This manuscript aligns with Kolmogorov's philosophical stance that information theory should not be subordinated to probability theory; rather, information theory is a more primordial bedrock, and — like Turing's edifice of computation — rests on logic. This ordering explains why Kolmogorov built algorithmic complexity on the Turing machine and then recovered Shannon entropy as the macroscopic expectation of microscopic Kolmogorov complexity. Conversely, the Brouwer-Heyting-Kolmogorov (BHK) interpretation of intuitionistic logic — in which the meaning of a proposition is identified with the **algorithmic method of its proof** — provides the logical bedrock of the **Curry-Howard Isomorphism**: **propositions as types, proofs as programs**. As civilization enters the AI Epoch, this treatise posits that its atomic substrate is the **TOKEN**. From this vantage point we advance a conjecture: just as the BIT forged a universal conduit between computing and communication, **the TOKEN will bridge the cognitive chasm between empiricism (associative memory, heuristic inference) and rationality (structural reasoning)** — or, in Kahneman's dichotomy, between **System 1** and **System 2** [149]. We thus arrive at the concluding synthesis: **whereas the BIT defined the Information Epoch, the TOKEN will define the generative architectures of the AI Epoch.**

Whether the current trajectory of LLMs will culminate in **Artificial General Intelligence** (AGI) or **Artificial Super Intelligence** (ASI) remains open. What this treatise maintains is that the AI Epoch has unmistakably begun, and that it is incumbent on the scientific community to reorganize foundational research around this newly crystallized atomic concept, constructing next-generation cognitive theories and generative systems accordingly. Whether the mechanistic act of predicting the next discrete token qualifies as "thinking" in the classical anthropocentric sense is debatable; what is not debatable is that autoregressive architectures have dramatically amplified humanity's capacity to synthesize, integrate, and process semantic information at scale. In the face of this philosophical frontier, we may reflect upon the provocation popularized in *The Imitation Game*: **The interesting question is, just because something thinks differently from you, does that mean it's not thinking?**

# Index of Figures